 \documentclass[preprint]{emulateapj}

\usepackage{epsfig}
\usepackage{graphicx}
\usepackage{amsmath}

\shorttitle{Probing interstellar dust with infrared echoes}
\shortauthors{Vogt et al.}

\begin{document}

\title{Probing interstellar dust with infrared echoes from the Cas A supernova}

\author{Fr\'ed\'eric P.A. Vogt\altaffilmark{1,2}, Marc-Andr\'e Besel\altaffilmark{2}, Oliver Krause\altaffilmark{2}, Cornelis P. Dullemond\altaffilmark{2,3}}
\email{fvogt@mso.anu.edu.au}

\altaffiltext{1}{Mount Stromlo Observatory, Research School of Astronomy and Astrophysics, The Australian National University, Cotter Road, Weston Creek, ACT 2611, Australia.}
\altaffiltext{2}{Max-Planck Institut f$\ddot{\textrm{u}}$r Astronomie, K$\ddot{\textrm{o}}$nigstuhl 17, 69117 Heidelberg, Germany}
\altaffiltext{3}{Zentrum f$\ddot{\textrm{u}}$r Astronomie, University of Heidelberg, Albert Ueberle Str.~2, 69120 Heidelberg, Germany}

\begin{abstract}
We present the analysis of an IRS 5-38~$\mu$m spectrum and MIPS photometric measurements of an infrared echo near the \object{Cassiopeia A} supernova remnant observed with the \emph{Spitzer Space Telescope}. We have modeled the recorded echo accounting for PAHs, quantum-heated carbon and silicate grains, as well as thermal carbon and silicate particles. Using the fact that optical light echo spectroscopy has established that Cas A originated from a type IIb supernova explosion showing an optical spectrum remarkably similar to the prototypical type IIb \object{SN~1993J}, we use the latter to construct template data input for our simulations. We are then able to reproduce the recorded infrared echo spectrum by combining the emission of dust heated by the UV burst produced at the shock breakout after the core-collapse and dust heated by optical light emitted near the visual maximum of the supernova light curve, where the UV burst and optical light curve characteristics are based on SN~1993J. We find a mean density of $\sim$680~H~cm$^{-3}$ for the echo region, with a size of a few light years across. We also find evidence of dust processing in the form of a lack of small PAHs with less than $\sim$300 carbon atoms, consistent with a scenario of PAHs destruction by the UV burst via photodissociation at the estimated distance of the echo region from Cas A. Furthermore, our simulations suggest that the weak 11~$\mu$m features of our recorded infrared echo spectrum are consistent with a strong dehydrogenated state of the PAHs. This exploratory study highlights the potential of investigating dust processing in the interstellar medium through infrared echoes. 
\end{abstract}


\section{Introduction}\label{Sec:intro}

Light echoes refer to light emitted by a bright source, such as a supernova (SN), which is scattered by dust located off the direct line-of-sight. The additional path length from the SN to the dust cloud and then to the Earth induces the time delay of the echo. Depending on the system considered and the brightness of the source, light echoes can be seen several hundreds of light years away from the source, and subsequently several hundred years after the initial SN outburst observation. 

Infrared (IR) echoes differ from optical light echoes in that the SN light is not scattered, but rather absorbed by dust, which will then re-radiate it at IR wavelengths. \cite{Bode80} considered this effect as the most plausible explanation for the IR excess seen in SN~1979c, a point which was later on confirmed by \cite{Dwek83}. IR echoes spanning over an area of several square-degrees on-sky have been discovered around the youngest known galactic core-collapse SN remnant Cassiopeia A (Cas A) \citep{Krause05}. \cite{Dwek08} used their spectra to deduce that these must have been created by a UV flash. The same echoes also enabled \cite{Kim08} to reconstruct a 3D map of the ISM surrounding Cas A, a concept that was previously used with SN~1987A by \cite{Xu99}.

The Cas A supernova remnant is believed to be the remains of a stripped-envelope type IIb core-collapse from a red supergiant progenitor with an initial mass in the range of 15-25~M$_{\odot}$ which might have lost much of its hydrogen envelope due to a binary interaction \citep{Young06,Krause08}, and is located at R.A.: 23$^{\mathrm{h}}$23$^{\mathrm{m}}$27.77$^{\mathrm{s}}$ ;  Dec: +58$^{\circ}$48'49.4" (J2000) \citep{Thorstensen01}. Studies of the proper motion of ejecta with the \emph{Hubble Space Telescope} suggest that the explosion occurred in 1681$\pm$19 \citep{Fesen06}. Sir John Flamsteed (1646-1719) might have visually observed the supernova on 1680 August 16, but it is not clear whether his mysterious \emph{3~Cassiopeia} observation was actually Cas A or not \citep{Ashworth80,Hughes80}. 

Along with the infrared echoes, several optical light echoes have been discovered around Cas A \citep[][]{Krause05,Krause08,Rest08}. \cite{Krause08} showed that the spectrum of one of them is strikingly similar to the spectrum of SN~1993J, which enabled the precise spectroscopic classification that Cas A was a type IIb SN. The knowledge of the exact type of the Cas A supernova will play a central role in our analysis.

SN~1993J exploded in M81 in late March 1993. It has been extremely well monitored across the entire spectrum from its very first moments \citep{Filippenko03}, including in the UV \citep[e.g.][]{Jeffery94}, the wavelength range at which the irradiation spectrum has critical consequences on interstellar dust. Such an amount of data makes it a target of choice for supernova simulations, that are able to simulate the observed spectrum already from its early phases \citep{Woosley94}. SN~1993J experienced an initial, short-lived (few hours) intense UV burst, later on followed by a longer (several days) less intense optical component \citep{Blinnikov98}. Hence, it is a very fortunate situation for our study that Cas A was of the same type and similar to SN~1993J - one of the brightest SNe in the northern sky for the last century, and for which a nearly unparalleled amount of data (including UV) exists.

Infrared echoes can be used to study the structure and composition of the thermally radiating dust, as was suggested by \cite{Dwek85} for circumstellar dust shells. Infrared echoes especially allow to study several dust processing mechanisms, such as photodissociation, dehydrogenation and ionization, in-situ. Due to the fact that more than 320 years have passed since Cas A exploded, the dust giving rise to the infrared echoes observed today is at least 160 light years away from Cas A. These echoes therefore most likely originate from interstellar rather than circumstellar dust. Such an analysis of infrared echoes is similar to what \cite{Sugerman03} suggested for optical light echoes, but using their infrared counterpart. As the SN light is absorbed and re-radiated at longer wavelengths, the resulting IR spectrum is a complex signature of both the SN light and the dust composition and characteristics. Using optical light echoes to identify the SN type, one can create a template light curve. Feeding this template in a suitable dust modelling program, one can simulate the IR echo spectrum - and use the resulting simulations to understand how ISM dust is behaving when illuminated by the SN light. 

One of the main advantages of this method is that IR light echoes potentially give us access to the consequences of \emph{pristine} interstellar material subject to a strong and short radiation burst. As such, assuming an undisturbed initial ISM, the dust signature detected in the IR light echo will be the consequence of the SN burst only, as no other mechanism, such as dust formation for example, can be acting on the very short timescale of a SN burst. 

In this article, we explore the potential of this approach by analyzing an IR echo spectrum recorded with the Spitzer Space Telescope around Cas A. We describe our observations and the resulting recorded IR echo spectrum in Sec.~\ref{Sec:obs}. We discuss how we simulate a light echo in Sec.~\ref{Sec:simu}, introduce our dust model in Sect.~\ref{Sec:simu_dust}, and present our Cas A SN template burst spectra in Sec.~\ref{Sec:SNbursts}. Our results are detailed in Sec.~\ref{Sec:results}. In Sec.~\ref{Sec:discussion}, we discuss how PAHs destruction (Sec.~\ref{Sec:depletion}), ionization (Sec.~\ref{Sec:ionization}) and dehydrogenation (Sec.~\ref{Sec:dehydro}) can improve our fit to the recorded IR echo spectrum. We compare our results with the work of \cite{Dwek08} in Sec.~\ref{Sec:dwek}. Finally, we explore the validity of our dust destruction hypothesis in Sec.~\ref{Sec:UV-link}, and present our conclusions in Sec.~\ref{Sec:conclusion}.

\section{Observations and data reduction}\label{Sec:obs}

\begin{figure*}[htb!]
\centerline{\includegraphics[scale=0.9]{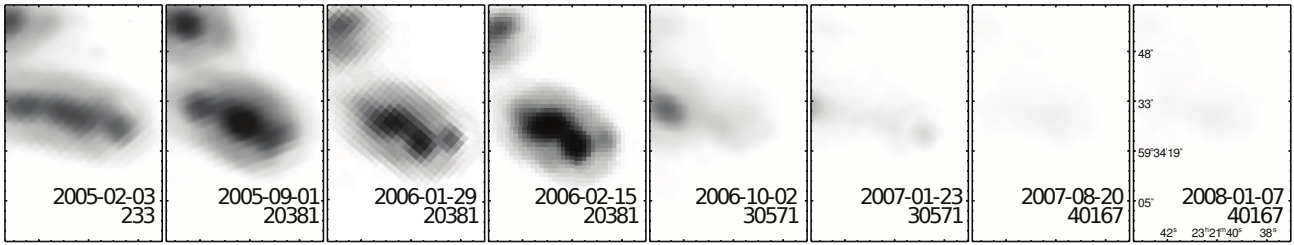}}
\caption{Time evolution of the echo region analyzed in this study. Observation date and Spitzer Program Identification (PID) are imprinted in each panel. All observations are MIPS 24$\mu$m images, except for an IRS red-peakup image (18.5-26~$\mu$m) on 2006 January 29. Each panel is displayed in logarithmic scale, with North up and East towards the left with a field of view of roughly $51''\times69''$.}\label{Fig:MA1}
\end{figure*}

The observations used in this study were obtained with instruments aboard the Spitzer Space Telescope \citep{Werner04}. After the first discovery of infrared echoes around Cas A by \cite{Hines04}, regular monitoring observations were carried out with the Multiband Imaging Photometer for Spitzer (MIPS; \cite{Rieke04}) (PIDs: 233, 20381, 30571; PIs: G.Rieke, O.Krause). A roughly 3~$\times$~3 square degrees field was observed every year and regions with especially interesting activity were observed at an interval of about half a year. MIPS observations were reduced using the MIPS Data Analysis Tool (DAT; \cite{Gordon05}), starting at the RAW data product level. 

Using those monitoring observations, a particularly bright source was detected at the location R.A.: 23$^{\mathrm{h}}$21$^{\mathrm{m}}$40$^{\mathrm{s}}$; Dec: +59$^{\circ}$ 34' 25''. Fig.~\ref{Fig:MA1} shows its evolution at 24$\mu$m over a time period of 3 years. It lies at an angular distance of 2857 arcsec away from the optical expansion center of Cas A. Although the source is already visible in the first epoch of observations, it changes its morphology, brightness and position drastically within just a few months time. In early 2006, it appears that most emission comes from three point like sources, suggesting that the real echoing structure is still unresolved with Spitzer. The emission diminishes very quickly between February and October 2006 indicating that the SN light beam swept past a probably very well defined end of the echo region.

\begin{figure}[htb!]
\centerline{\includegraphics[scale=0.4]{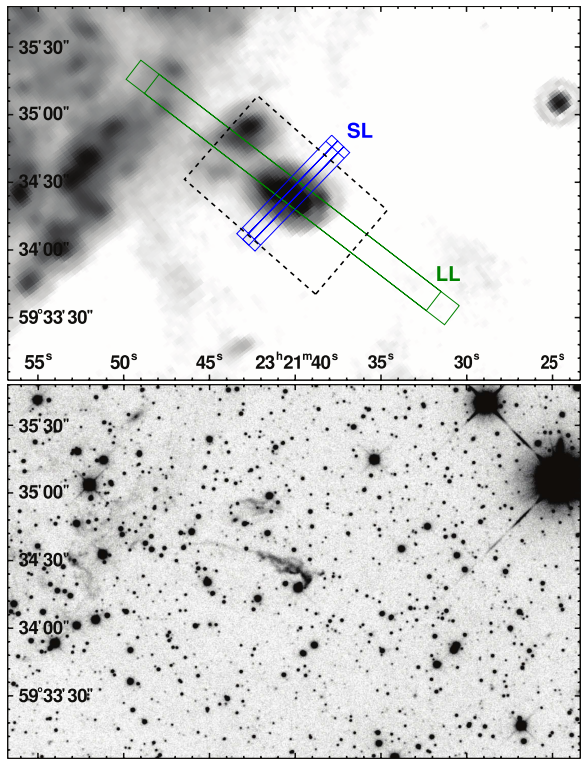}}
\caption{Top: 24~$\mu$m aerial image of the echoing source with IRS red-peakup insert (dashed rectangle) and footprints of the Long-Low (LL, in green in the online version) and Short-Low (SL, in blue in the online version) spectroscopic slits positions. Bottom: Calar Alto Omega 2000 Ks-Band image of the same area. Visible are tiny structures within the infrared emitting area.}\label{Fig:MA3}
\end{figure}

On 2006 January 29, a low-resolution spectrum of this bright echoing source was obtained with the Infrared Spectrograph (IRS; \cite{Houck04}) as part of program (PID) 20381 (P.I.: O. Krause). A 24~$\mu$m aerial image of the echoing source is shown in Fig.~\ref{Fig:MA3}, where we have overlaid the IRS red peak-up insert (rectangle) as well as the spectroscopic slits position; 1$\times$ Long-Low (LL) and 3$\times$ Short-Low (SL). The red peak-up functionality of IRS was used to achieve the best possible centering of the slits on this fast moving echo-feature. In addition to space-based infrared observations, we also obtained ground based near-infrared ($\lambda$=2.2$\mu$m) Ks-band images for this echo region using Omega 2000 on the Calar Alto (Spain) 3.5m telescope in December 2005, which is shown in the bottom of Fig.~\ref{Fig:MA3} for comparison. These images show the similar overall structure of the region, but reveal very small scale filamentary substructure of the echo region. A full analysis of these observations is to be presented in Besel et al. (in preparation). It should be noted here that the Ks-band as well as the J-band and H-band images (which have similar morphology, and are not shown here) show the scattered light echo, not the IR radiating dust echo.

The S18.7.0 version of the Basic Calibrated Data (BCD) products from the Spitzer Science Center (SSC) pipeline were fed into SMART \citep{Higdon04,Lebouteiller10} for final reduction. Depending on the used module, total exposure times ranged between 120s and 300s. One spectra was extracted for each of the two brightest point-like sources. They were averaged separately and combined later. A background was estimated based on the non-echo-contaminated parts of either the off-source order or the individual slits (contamination is easily evident on the MIPS 24~$\mu$m image for one end of the LL slit). As the SL and LL modules cover slightly different special regions of the echoing source the resulting spectra were scaled to match each other's continuum levels. Absolute calibration was verified by comparing the integrated IRS flux and the measured 24~$\mu$m photometry. 

\begin{figure}[htb!]
\centerline{\includegraphics[scale=0.45]{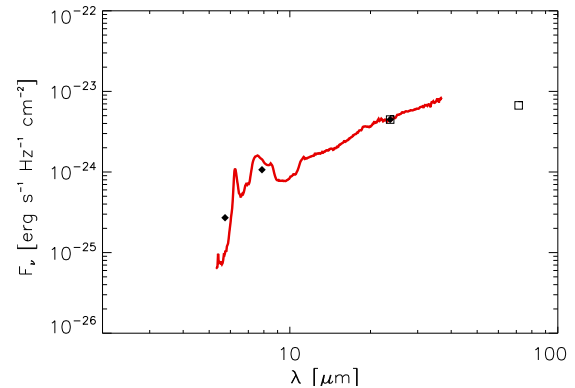}}
\caption{Final extracted IRS spectrum of the echo region (in red in the online version) and associated Spitzer synthetic photometric measurements (black rombuses). The white squares correspond to MIPS 24~$\mu$m and 70~$\mu$m photometric measurements of the same echo region. The spectrum and synthetic photometric points have been scaled to match the 24~$\mu$m MIPS photometry of the entire region.}\label{Fig:MA4}
\end{figure}

Fig.~\ref{Fig:MA4} shows the final combined extracted spectrum of this IR echo. Also plotted are MIPS 24~$\mu$m and 70~$\mu$m photometric measurements within a 22.2~arcsec aperture (white squares). Note that their error bars are of the order of the symbol size.  Black rombuses at 5.7~$\mu$m, 7.9~$\mu$m and 24~$\mu$m correspond to the Spitzer Synthetic Photometric points of the echo spectrum, reconstructed using the Recipe 10 of the Spitzer Data Analysis Cookbook \footnote{see http://ssc.spitzer.caltech.edu/dataanalysistools/cookbook/}. As the IRS slits do not cover the entire echo region, the spectrum and synthetic photometric points were scaled to match the 24~$\mu$m photometry of the entire region. 

\section{Simulating the infrared echo}\label{Sec:simu}

The spectrum and photometric points presented in Fig.~\ref{Fig:MA4} are the result of the interaction of the Cas A SN burst and a region of the ISM located away from the SN center. In that respect, the data potentially contain information on both the SN explosion and the echoing dust. A simple visual inspection of the spectrum already reveals a very weak 11~$\mu$m complex of PAH features as compared to the strength of the 7~$\mu$m PAH features, a point that will be discussed further in Sec.~\ref{Sec:dehydro}. 

Here, we adopt the following strategy to study and extract the underlaying information on the echoing dust contained in the measured echo: we implement a dust modelling program (see Sect.~\ref{Sec:simu_dust}) to simulate the IR response of ISM dust when heated by radiation with a certain spectral energy distribution (SED). To obtain the dust thermal emission response, its distribution along the line of sight has to be convolved with the corresponding irradiation spectrum from the supernova. We assume that the effective radiation field can be decomposed by (I) the UV dominated SED during the first hours after the shock outbreak and (II) the optically dominated SED at the epoch near maximum light (see Sect.~\ref{Sec:SNbursts}). We adopt those two template SEDs to account for the convolution of the input radiation spectrum with the echo region on length scales of at least a few light days.

\begin{figure}[htb!]
\centerline{\includegraphics[scale=0.5]{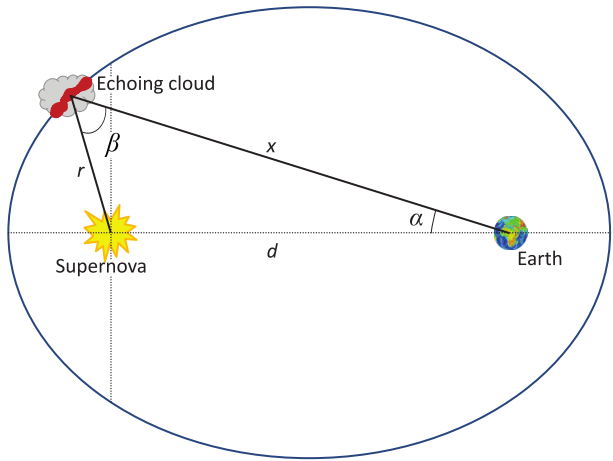}}
\caption{Light-Echo geometry. The observer on Earth and the Supernova are located in the foci of an ellipsoid. All points on this ellipsoid describe locations with equal delay time t which corresponds to the time difference between the original SN outburst observation and the current observing epoch. Material situated on any point of the ellipsoid can simultaneously be seen as an echoing source.}\label{Fig:MA2}
\end{figure}

In Fig.~\ref{Fig:MA2}, we present a geometrical sketch of the echo principle. The position of the echo region is located, at any given time, on a ellipsoid with Cas A and the Earth at the foci \citep{Couderc39}. The distance $r$ between the echo region and the center of Cas A is :
\begin{equation}\label{Eq:r}
r=\frac{d(d+ct)\big(1-\cos(\alpha)\big)+(ct)^2/2}{d\big(1-\cos(\alpha)\big)+ct}
\end{equation}
where $t$ is the time delay between the light of Cas A's explosion reaching Earth and the observation of the light echo, $d$ the distance to Cas A and $\alpha$ the on-sky angular separation between Cas A and the echo region. Using a distance to Cas A of 3.4 kpc \citep{Reed95} and an explosion date of AD 1681 \citep{Fesen06}, the absolute distance of the echo region to the supernova remnant is $r=199$ light years.

\begin{figure}[htb!]
\centerline{\includegraphics[scale=0.4]{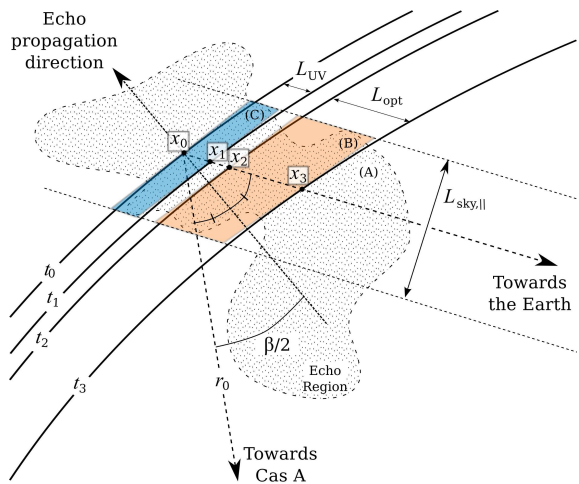}}
\caption{Schematic of the echo region. The echo emission is coming from two defined regions (B) and (C) heated by the optical and UV burst respectively. The drawing depicts \emph{what we see at a given time} from the Earth, which is different from the \emph{state of the echo region at a given time}. With a thickness of a few light days, the excited slabs are much smaller than the actual echo region, which is not completely excited at one given moment in time (region (A)).}\label{fig:drawing}
\end{figure}

A schematic of the echo region itself is shown in Fig.~\ref{fig:drawing}. Because the UV burst and optical phase do not occur simultaneously, they will heat a given area of the echo region sequentially. In Fig.~\ref{fig:drawing}, area (A) corresponds to the echo region not seen excited at the moment of the observation, while areas (B) and (C) denote respectively the schematic location of the UV burst and optical light curve excited slabs, moving away from Cas A. 

It is important to note here that if we \emph{see} those excited slabs simultaneously, they \emph{are not} excited simultaneously, because of the finite light speed. The UV-burst excited slab (a.k.a. UV-heated slab) is located farther away from Cas A and the Earth, and must therefore be excited earlier than the optically-excited slab (a.k.a optically-heated slab) to be seen at the same time on Earth. In situ, the two SEDs will be seen by a dust grain with the same time interval as when they occurred after the shock breakout. As seen from the Earth, the excited slabs are encompassed in between ellipsoids of different radii $r_0$, $r_1$, $r_2$ and $r_3$ defined by Eq.~\ref{Eq:r}, with $t_0$, $t_1$, $t_2$ and $t_3$ the delay times between the observation and the shock breakout, the end of the UV burst, the start of the optical phase and the end of the optical phase, respectively.

Let us define $L_{\mathrm{sky},\parallel}$ and $L_{\mathrm{sky},\perp}$ the on-sky dimensions of the echo, in the direction of and perpendicular to Cas A, respectively. Because the UV-heated and optically-heated slabs thickness will only be of a few light days, the excited volume of the echo region \emph{ visible at one given moment in time} is much smaller than the total volume of the echo region. For simplification, because we do not resolve individual ISM structures and because the echo region is small compared to the ellipsoid and its local curvature depicted in Fig.~\ref{Fig:MA2}, let us assume the excited slabs to be parallelepipeds, of which the sections are depicted by the light-grey (B) and dark-grey (C) regions in Fig.~\ref{fig:drawing} (in orange and blue in the online version). Their height (perpendicular to the drawing plane) is $L_{\mathrm{sky},\perp}$. Their respective thickness $L_{\mathrm{UV}}$ and $L_{\mathrm{opt}}$ along the line-of-sight are :
\begin{eqnarray}
L_{\mathrm{UV}}=(x_0-x_1)\label{eq:2}\\
L_{\mathrm{opt}}=(x_2-x_3)\label{eq:3}
\end{eqnarray}
where, from Fig.~\ref{Fig:MA2} and the law of cosines :
\begin{eqnarray}
x_i&=&ct_i+d-r_i\qquad\forall\mathrm{ }i=1,2,3,4\label{Eq:x0}
\end{eqnarray}
For the echo considered here and presented in Sec.~\ref{Sec:obs}, $x_0\cong3439$~pc. We can now define an upper limit to the \emph{visible} volume of the echo region, $V_{\mathrm{echo,max}}$, as seen from the Earth at any given time :
\begin{equation}\label{eq:volume}
V_{\textrm{echo,max}}=L_{\mathrm{sky},\parallel}\times (L_{\mathrm{UV}} + L_{\mathrm{opt}})\times L_{\mathrm{sky},\perp}
\end{equation}
where $L_{\mathrm{sky,\parallel}}\times L_{\mathrm{opt}}$ and $L_{\mathrm{sky,\parallel}}\times L_{\mathrm{UV}}$ are the respective areas of the regions (B) and (C) in Fig.~\ref{fig:drawing}. The shape of a fictional echo region (dotted surface) depicted in Fig.~\ref{fig:drawing} illustrate why $V_{\mathrm{echo,max}}$ is an upper limit on the visible volume of the echo region, as the dust does not, in principle, have to fill completely the areas (B) and (C). We delay the numerical computation of $V_{\mathrm{echo,max}}$ until Sec.~\ref{sec:fit}, after introducing our adopted template UV burst and optical light curve in Sec.~\ref{Sec:SNbursts}.

\section{Simulating ISM dust}\label{Sec:simu_dust}
Simulations provide an efficient tool to study the theoretical behaviour of interstellar dust in an attempt to unveil its composition and detailed characteristics. Various models, such as the silicate core-carbonaceous \citep[][]{Desert90,Jones90,Li97}, the composite \citep[][]{Mathis89,Mathis96,Zubko04} or the silicate-graphite-Polycyclic Aromatic Hydrocarbons (PAHs) model \citep[][]{Draine84,Siebenmorgen92,Li01,Draine07} have been proposed to reproduce the behaviour of ISM dust. To study the recorded IR echo spectrum shown in Fig.~\ref{Fig:MA4}, we have implemented our own modelisation program, strongly based on the dust model of \cite{Li01}. Our dust mix contains PAHs, quantum-heated carbon and silicate grains, as well as thermal graphite and silicate grains. Similarly to \cite{Li01}, the PAHs and quantum-heated carbon grains are mixed and form what is referred to as the quantum-heated carbonaceous dust. The \cite{Li01} model is a well established and observationally tested model for the average dust composition of the general Milky Way interstellar medium. Since the echo region considered in this study is located in the general ISM at a distance of $\sim$199 light years from the supernova, using this model appears very appropriate.

We simulate particles from $\sim$3.5~{\AA} up to $\sim$6000~{\AA}, and have a transition from quantum-heated to thermal grains at $\sim$250 \AA. The dust mix is following the formula of \cite{Weingartner01} multiplied by 0.92, as suggested by \cite{Draine07}, from whom we have implemented the updated Drude profiles for the PAHs opacities. Our simulation yet differs from the \cite{Li01} model in a few aspects. We use enthalpies from \cite{Siebenmorgen92}, which are based on a formula by \cite{Chase85}. Our silicate opacities are calculated via the Mie theory using scattering and absorption cross section from \cite{Ivezic97}. Finally, we also use enthalpies of graphite for the silicate grains for simplification.

Quantum-heated grains are subject to stochastic heating by the incoming photons. They do not behave as constant-temperature grains, but experience strong temperature variations, and a careful approach is required to compute their emission. More precisely, their emissivity is obtained by calculating the temperature probability distribution for a certain grain size, and we refer the reader to the work of \cite{Draine01,Li01,Draine07} for detailed explanations regarding the computation of the emissivity of small grains. Large thermal grains, on the other hand, will experience much smaller temperature excursions, and can be approximated as being in thermal equilibrium. Under this assumption, their emissivity is obtained by comparing the incoming radiation to the re-radiated energy \citep{Li01} :
\begin{equation}
\int_0^{\infty}C_{\mathrm{abs}}(a,\lambda)F_{\lambda}(\lambda)d\lambda=\int_{0}^{\infty}C_{\mathrm{abs}}(a,\lambda)4\pi B_{\lambda}(\bar{T})d\lambda
\end{equation}
with $F_{\lambda}(\lambda)$ the incoming radiation flux at wavelength $\lambda$, $\bar{T}$ the equilibrium mean temperature of the grain, $B_{\lambda}(\bar{T})$ the Planck function and $C_{\mathrm{abs}}(a,\lambda)$ the absorption cross-section of a grain of radius $a$ at the wavelength $\lambda$.

\begin{figure}[htb!]
\centerline{\includegraphics[scale=0.5]{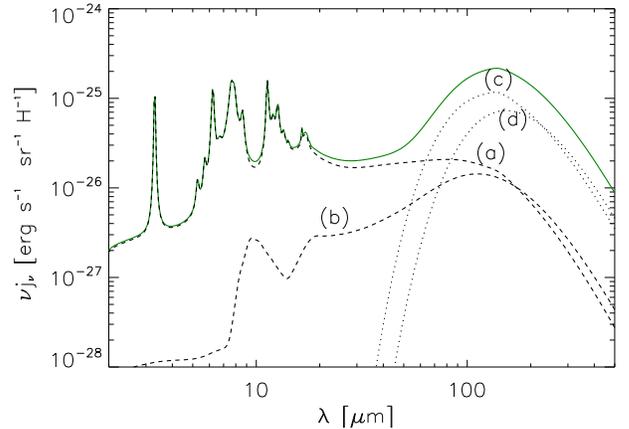}}
\caption{IR emissivity of interstellar dust per H nucleon heated by the ISRF of \cite{Mathis83}, assuming a silicate-graphite-PAHs composition.  Dashed and dotted curves show the individual contribution of the quantum-heated carbonaceous dust (a), quantum-heated silicate dust (b), thermal graphite (c) and thermal silicate grains (d). }\label{Fig:Li01-HGL}
\end{figure}

In Fig.~\ref{Fig:Li01-HGL}, we present our simulation (full line, in green in the online version) of the ISM dust IR emissivity when heated by the standard Interstellar radiation field (ISRF) of \cite{Mathis83} (see Fig.~\ref{Fig:bursts}). The contribution of each individual dust components is plotted in dashed (quantum-heated grains) and dotted (thermal grains) lines; (a) - quantum-heated carbonaceous dust, (b) - quantum-heated silicate dust, (c) - thermal graphite grains, (d) - thermal silicates grains.  The simulation code returns the emissivity of the dust mix per H nucleon, $j_{\nu}$, in units of erg s$^{-1}$ sr$^{-1}$ H$^{-1}$ Hz$^{-1}$, and in Fig.~\ref{Fig:Li01-HGL}, we plot the spectrum in the form of $\nu j_{\nu}$, in units of erg s$^{-1}$ sr$^{-1}$ H$^{-1}$.

\begin{figure}[htb!]
\centerline{\includegraphics[scale=0.38]{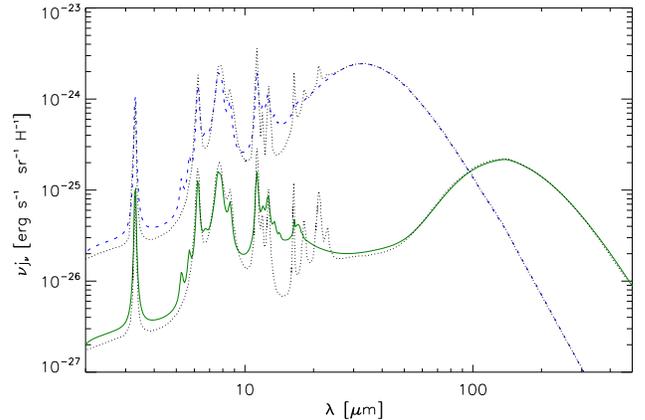}}
\caption{IR emissivity of interstellar dust per H nucleon for HGL dust (full line, in green in the online version) and for ISM dust illuminated by 10$^{4}$ times the ISRF (dashed line, in blue in the online version). The latter has been scaled down by 10$^{3}$. The corresponding simulations of \cite{Li01} are shown in dotted lines.}\label{Fig:Li01-chi}
\end{figure}

We validate our code output against the results of the \cite{Li01} code, and the comparison is made in Fig.~\ref{Fig:Li01-chi}. Our HGL simulation (full line, in green in the online version) reproduces closely the one from \cite{Li01} (dotted line), with the exception of the PAH features, that have been updated to reproduce the simulations of \cite{Draine07}. Note that the spectrum of \cite{Li01} has been scaled by 0.92 to allow for comparison. 

The strategy to simulate the IR echo spectrum presented in Sec.~\ref{Sec:obs} is to use our code with a template input spectrum, replacing the ISRF of \cite{Mathis83} by a theoretical SN burst spectrum. In order to validate that our code is stable with a more intense heating radiation field, we simulate ISM dust heated by a radiation field equal to 10$^{4}$ times the ISRF, and compare our result with the similar simulation of \cite{Li01} (see Fig.~13 in their article). The comparison is shown in Fig.~\ref{Fig:Li01-chi} (dashed line, in blue in the online version). It is clear that we reproduce very closely the \cite{Li01} simulations for stronger radiation fields, with exception to, once again, the intrinsic differences due to the update of the Drude profiles for the PAHs, which match the simulations of \cite{Draine07}. We conclude that our simulation code is stable to intensity variations of the incoming radiation field.

\section{Modelling the Cas A SN outburst}\label{Sec:SNbursts}

Simulating the spectrum of an infrared echo requires the knowledge of the exact radiation SED heating the dust. Generally, this is not known with precision, as the SN light swept past Earth many years before a light echo observation. Fortunately, as we mentioned in Sect.~\ref{Sec:intro}, SN~1993J has been found to be a very good template for Cas A, and has been the subject of many observations and simulations over a large range of wavelengths. Hence, following the characteristics of SN~1993J, we make the following simplification mentioned previously: we consider the Cas A SN explosion to be made of two distinct phases - a UV burst and an optical light phase. Each phase consists of a single SED, which we take to be representative of the actual spectrum variations over the phase duration. 

To create the template UV burst spectrum, we use the simulations of \citet{Blinnikov98}. In this model, the UV burst is close from its maximum luminosity during a few hours. The total UV burst lasts a few days, with its luminosity reaching a minimum after about $\sim$10 days. Energetic UV photons have a strong effect on dust and especially on PAHs, as we will discuss later on. It is therefore critical to have a representative SED that contains those energetic UV photons. The peak of the UV burst is however fairly brief, and one has to pay attention not to over-estimate the amount of very energetic photons. With these considerations in mind, we adopt our template Cas A UV-burst to be a black-body spectrum, with a temperature of  $T_{\mathrm{uv}}=1.5\times10^{5}$~K. In an attempt to provide a slightly more realistic light curve, we subsequently adapt the shape of the black-body curve to that of the .278 curve of \cite{Blinnikov98} in the 0.001 - 1~$\mu$m range. We assume a burst duration of 2.5 days, and a total luminosity of L$_{\mathrm{uv-B}}=6.4\times10^{44}$~erg~s$^{-1}$. For comparison, the luminosity of the black body curve prior to the shape modification is L$_{\mathrm{uv}}$=5.9$\times$10$^{44}$~erg~s$^{-1}$. By modifying the burst shape rather than increasing the burst temperature, we ensure that we do not overestimate the number of very energetic photons, while still taking into account the fact that a significant amount of energy is radiated away during a short period of time. To account for the smaller, but yet non-negligible, burst luminosity beyond the 2-days mark, we adopt a total burst duration of 2.5 days, shorter than the $\sim10$ days mark, to avoid an obvious over-estimate.

We concede that there exist many other ways to obtain a Cas A template UV burst from the data of SN~1993J, but any solution, including our adopted template, would involve quite a number of assumptions. We emphasize here that we do not claim to have implemented neither the best, nor the most accurate reproduction of Cas A's UV-burst. But rather made an good educated guess. In Section~\ref{Sec:dwek}, we will compare our UV burst template with previous estimates by \cite{Dwek08} and show that both UV burst estimates are consistent.

We use the work of \citet{Richmond94} to create the template optical light curve. They find the optical phase to last some 20-30 days, with a luminosity depending on the color excess E(B-V) on the line of sight. They provide two estimates, a low one with E(B-V)~=~0.08, and a moderate one, with E(B-V)=0.32. Based on the work of \cite{Clocchiatti95}, that finds E(B-V)=0.25~$\pm$~0.05, and using the discussion on the absorption on the line-of-sight towards SN~1993J of \cite{Matheson00}, that adopt in their work a value of E(B-V)~=~0.19, we assume our optical light curve to have the characteristics of the \cite{Richmond94} estimates (see Table~15 in their paper) adapted for a color excess E(B-V)$\cong$0.20. Our optical light curve is a black-body spectrum, with a temperature $T_{\mathrm{opt}}=9.4\times10^3$~K (between the $\sim7.9\times10^3$ and $\sim11.0\times10^3$ estimates of \cite{Richmond94}), a luminosity of $L_{\mathrm{opt}}=3.8\times10^{42}$~erg~s$^{-1}$ (between the $\sim1.7\times10^{42}$ and $\sim7.1\times10^{42}$ estimates of \cite{Richmond94}) and a duration of 25 days (see Fig.~10 in \cite{Richmond94}). We stress again that we consider this optical light curve as a good educated guess rather than the perfect estimate.

\begin{figure}[htb!]
\centerline{\includegraphics[scale=0.37]{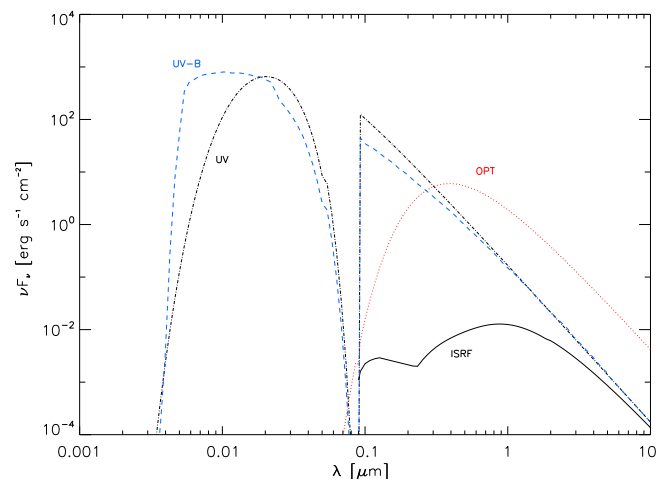}}
\caption{The UV-B (dashed line, in blue in the online version) and optical (dotted line, in red in the online version) light curve used to simulate the recorded IR echo spectrum shown in Fig.~\ref{Fig:MA4}. The UV light curve (dot-dashed) is a black body with T=1.5$\times$10$^{5}$~K and is shown for comparison, along with the ISRF curve (full line) from \cite{Mathis83}.}\label{Fig:bursts}
\end{figure}

The resulting UV burst and optical light curve fluxes $\nu F_{\nu}$ shining on the dust in units of erg~s$^{-1}$~cm$^{-2}$ are shown in Fig.~\ref{Fig:bursts}. The final UV burst spectrum used in our simulations, of which the shape has been updated based on the work of \cite{Blinnikov98}, is labelled 'UV-B' (dashed line), while the black body UV burst with T=1.5$\times$10$^{5}$~K is labelled 'UV' (dot-dashed line) and is plotted for comparison. Using the IDL\footnote{Interactive Data Language} ismtau.pro\footnote{see, e.g. http://hea-www.harvard.edu/PINTofALE/pro/ismtau.pro (accessed on 2011 December 11)} routine, we have taken into account some foreground absorption for the UV burst assuming an atomic H--column density $N_H$ of $10^{18}$ cm$^{-2}$, an H~II / H~I ratio of 0.26, an He~I~-- and He~II~--~column densities of $0.1~N_H$, and used the photoionization cross-section of \cite{Morrison83}. Note that we have not taken into account any intrinsic extinction by the echo region itself. The various burst characteristics are summarized in Table~\ref{Table:bursts}.

\begin{table*}[htb!]
\begin{center}\caption{The template UV burst and optical light curve characteristics used in our simulations.}\label{Table:bursts}

\begin{tabular}{c c c c c c}
&&&&\\
\tableline
\tableline
Burst type& T$_{Wien}$ [K]& Luminosity~[erg s$^{-1}$] & Absorption & Duration [days] & Total Energy [erg]\\
\tableline
Optical&$9.4\times10^{3}$&$3.8\times10^{42}$&No & 25 & $8.2\times10^{48}$\\
UV-B&-&$6.4\times10^{44}$&Yes & 2.5 & $1.4\times10^{50}$\\
\hline
UV&$1.5\times10^{5}$&$5.9\times10^{44}$&Yes & - & -\\
\tableline
\end{tabular}
\end{center}
\end{table*}

\section{Results}\label{Sec:results}

\subsection{Fitting the echo spectrum}\label{sec:fit}
With the adopted UV burst and optical light curve characteristics defined in the previous section, we can now get the estimates for the size of the echo region. Specifically, with $t_0=325$~year, $t_0-t_1=2.5$ days, $t_1-t_2=10$ days, $t_2-t_3=25$ days, $\alpha=2857$~arcsec and $L_{\mathrm{sky},\parallel}=L_{\mathrm{sky},\perp}\cong 1.2 $ light years (22" on-sky aperture, see Sec.~\ref{Sec:obs}), we get, using Eq.~\ref{eq:2} to Eq.~\ref{eq:volume} :
\begin{eqnarray}
L_{\mathrm{UV}}&\cong&1.53 \textrm{ light days}\nonumber\\
L_{\mathrm{opt}}&\cong&15.3 \textrm{ light days} \nonumber\\
V_{\mathrm{echo,max}}&\cong&6.4\times10^{-2} \textrm{ (light years)}^{3} \label{Eq:volume}\\
&\cong&5.5\times10^{52}\textrm{ cm}^3\nonumber
\end{eqnarray}
Because we do not resolve the echo region, we make the assumption that $F_{\mathrm{echo}}$ the recorded echo flux is equal to the sum of the UV-heated slab flux and the optically-heated slab flux $F_{\mathrm{UV}}$ and $F_{\mathrm{opt}}$ respectively :
\begin{equation}\label{eq:flux}
F_{\textrm{echo}}=F_{\textrm{UV}}+F_{\textrm{opt}}
\end{equation}
Furthermore, $F_{\mathrm{UV}}$ and $F_{opt}$ relate to the UV-heated and optically-heated dust emissivities $j_{\nu,\mathrm{UV}}$ and $j_{\nu,\mathrm{opt}} $ in the following way :
\begin{eqnarray}
F_{\mathrm{UV}}= j_{\nu\mathrm{,UV}}\times\frac{M_{\mathrm{dust,UV}}}{m_{\mathrm{H,dust}}\cdot x^2}\\
F_{\mathrm{opt}}=j_{\nu\mathrm{,opt}}\times\frac{M_{\mathrm{dust,opt}}}{m_{\mathrm{H,dust}}\cdot x^2}
\end{eqnarray}
where $M_{\mathrm{dust,UV}}$ and $M_{\mathrm{dust,opt}}$ are the echoing dust mass, in g, heated by the UV burst and the optical light curve respectively, $m_{\mathrm{H,dust}}$ is the dust mass per H nucleon in g~H$^{-1}$, and $x$ is the distance from the Earth to the echo region in cm. We assume $m_{\mathrm{H,dust}}=1.89\times10^{-26}$ g~H$^{-1}$ \citep{Li01}. From our assumptions in Sec.~\ref{Sec:simu} (see also Fig.~\ref{fig:drawing}), we can express $M_{\mathrm{dust,UV}}$ and $M_{\mathrm{dust,opt}}$ as a function of the cloud density $\rho_{\mathrm{cloud}}$, and the volume of the excited slabs (see Eq.~\ref{eq:2} and \ref{eq:3}) :
\begin{eqnarray}
M_{\mathrm{dust,UV}}=  \rho_{\mathrm{echo}}\times L_{\mathrm{sky},\perp}\times L_{\mathrm{sky},\parallel}\times(x_0-x_1)\label{eq:dopt0}\\
M_{\mathrm{dust,opt}}=  \rho_{\mathrm{echo}}\times S_{\mathrm{sky},\perp}\times L_{\mathrm{sky},\parallel}\times(x_2-x_3)\label{eq:dopt}
\end{eqnarray}
Hence, using Eq.~\ref{eq:flux} to \ref{eq:dopt}, it is possible to express the echo spectrum $F_{\mathrm{echo}}$ as a function of the total mass of echoing dust $M_{\mathrm{dust}} = M_{\mathrm{dust,UV}} + M_{\mathrm{dust,opt}}$ in the following way : 
\begin{equation}\label{Eq:fit}
F_{\mathrm{echo}}=\left[\zeta\cdot j_{\nu\mathrm{,UV}}+(1-\zeta)\cdot j_{\nu\mathrm{,opt}}\right]\times\frac{M_{\mathrm{dust}}}{m_{\mathrm{H,dust}}\cdot x^2}
\end{equation}
where $M_{\mathrm{dust}}$ is in g and $\zeta=\frac{M_{\mathrm{dust,UV}}}{M_{\mathrm{dust}}}$ is the ratio of the mass of the UV-slab dust to the total echoing dust mass, which, using Eq.~\ref{eq:dopt0} and \ref{eq:dopt}, can be written as :
\begin{equation}\label{Eq:zeta}
\zeta=\frac{x_0-x_1}{(x_0-x_1) + (x_2-x_3)}=0.09
\end{equation}
Eq.~\ref{Eq:fit} and \ref{Eq:zeta} reflect the fact that the UV-heated and optically-heated dust emissivities \emph{are mixed} with a 9\%/91\% ratio (similar to the UV vs optical phase duration ratio). For simplification and clarity, we will refer to this ratio as our \emph{9\%/91\% e-mix} (for emissivity mix) in the rest of this article. We emphasize however that this is not equal to the \emph{luminosity} ratio of the UV-heated to optically-heated component of the simulated echo spectrum, which we will show in the subsequent Sections is of the order of $\sim$80\%/20\% in the 5-38 $\mu$m range. 

We also note that this 9\%/91\% ratio is a direct consequence of our assumptions to have a symmetric echo region. Rapid size variations would lead to an alteration of the mixing ratio, and potentially rapid variability of the light echo. With only one echo spectrum, we are not able to address these questions, which will require the monitoring of one given echo region over a relatively short period of time.

Ultimately, in Eq.~\ref{Eq:fit}, $M_{\mathrm{dust}}$ is the only undefined parameter, and we use it as the fitting parameter between the recorded spectrum and our simulated one (i.e. we use it to move the simulated spectrum vertically to match the intensity of the recorded echo spectrum).

\subsection{Simulation results}

\begin{figure}[htb!]
\centerline{\includegraphics[scale=0.37]{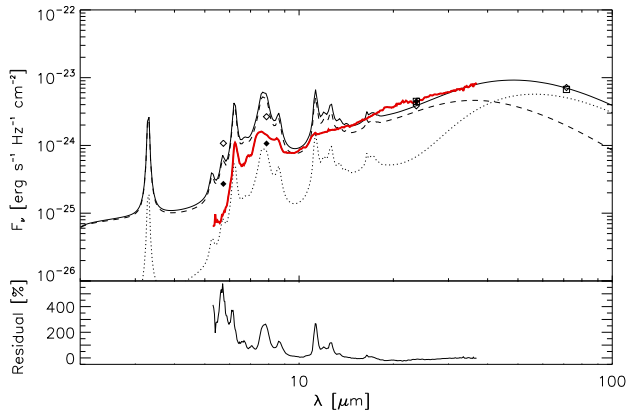}}
\caption{Simulation SIM1. Top : 9\%/91\% e-mix simulation (thin line) fitted to the recorded IR echo (thick line, in red in the online version) shown in Fig.~\ref{Fig:MA4}, and associated synthetic photometric points (black rhombuses). The white squares correspond to MIPS 24~$\mu$m and 70~$\mu$m photometric measurements. The white rhombuses represent the Spitzer synthetic photometric measurements for our simulation. The dashed and dotted lines correspond to the individual emission from UV-heated and optically heated dust, respectively. Bottom : fit residuals relative to the recorded echo spectrum.}\label{Fig:fit-1}
\end{figure}

The resulting simulated spectrum (labelled SIM1), using standard ISM dust (described in Sec.~\ref{Sec:simu_dust}) and a 9\%/91\% e-mix (described in Sec.~\ref{sec:fit}), is shown in Fig.~\ref{Fig:fit-1} (top panel). Fitting the intensity of the simulated spectrum to that of the recorded one (thick line) using Eq.~\ref{Eq:fit} leads to an echoing dust mass of $6.6\times10^{29}$~g, or $3.3\times10^{-4}$~M$_{\sun}$. The dashed and dotted lines correspond to the individual contributions from UV-heated and optically-heated dust, respectively. In the 5-38 $\mu$m range, the UV component of the simulated echo spectrum accounts for 80.0\% of the total luminosity of the simulated echo. Quite interestingly, the optical component influence becomes more important than that of the UV one beyond 50 $\mu$m. The presence of an optical component especially appears essential in order to fit the 70$\mu$m photometric measurement. It will therefore be of critical importance to obtain such photometric measurements beyond 50$\mu$m in future studies of infrared echoes if one intends to better characterize the optically-heated dust component of such echoes.

Given the total echoing dust volume found in Eq.~\ref{Eq:volume}, the resulting echo region density is $\rho_{\mathrm{echo}}=640$ H~cm$^{-3}$ or $1.2\times10^{-23}$~g~cm$^{-3}$, with a column density of $2.8\times10^{19}$~cm$^{-2}$. Because the volume of echoing dust calculated in Sec.~\ref{sec:fit} is an upper limit of the actual volume of echoing dust (see Sec.~\ref{Sec:simu}), the value of $\rho_{\mathrm{echo}}$ can be considered as a lower limit on the actual echoing dust density.

The bottom panel shows the fit residual (relative to the recorded echo). The general trend of the recorded spectrum is well reproduced above $\sim$15~$\mu$m, as we match both the recorded echo spectrum as well as the 70~$\mu$m photometric point. Yet, there are some discrepancies in the PAHs features, where our simulated spectrum is significantly too strong. We especially fail, at this point, to account for the peculiarly small 11~$\mu$m complex. The discrepancies around the PAHs features is reflected in the difference between the recorded and simulated photometric points shown in the bottom panel of Fig.~\ref{Fig:fit-1}, with a $\sim$250\% overestimate of the flux in those regions. In the following Section, we modify our assumptions on the echoing dust to improve the fit below 15~$\mu$m.

\section{Discussion}\label{Sec:discussion}

\subsection{On the influence of the dust model}

As we base our simulations on the dust model of \cite{Li01,Draine07}, it is clear that resulting dust masses and densities of the echoing region are a direct consequence of this model. Alternate dust models, such as the ones mention in Sec.~\ref{Sec:simu_dust},  are based on different total abundances (which can vary along different line-of-sights), dust characteristics, mixtures and dust-to-gas ratios. Because the aim of this paper is \emph{not} to compare the \emph{goodness} of different dust models and their ability to reproduce our recorded echo spectrum, but rather to show the potential of performing such a fit and its subsequent scientific output, we deliberately choose to alter the original dust model of \cite{Li01,Draine07} \emph{only} to explore the potential of such modifications to improve our fit to the recorded echo spectrum, but \emph{not} in an attempt to solely improve their dust model. Performing a similar analysis to the one presented in this article, but with a different dust model, has a great potential but is outside the scope of this paper.

\subsection{The 3-20~$\mu$m features: PAHs depletion}\label{Sec:depletion}

PAHs can be strongly affected by UV radiation, which can cause photoionization, dehydrogenation or photodissociation \citep{Tielens08}. If those various processes are not directly implemented in our simulation code, we can nevertheless (approximatively) take them into account by varying the amount of various dust components in the code, so as to reproduce their effect on the dust composition. Specifically, we modify the equation responsible for the mix in between PAHs and quantum-heated carbon grains as follows (see Eq.~2 in \cite{Li01}): 
\begin{equation}\label{Eq:xi}
C_{\mathrm{abs}}^{\mathrm{carb}}(a,\lambda)=\eta(a)\xi_{\mathrm{PAH}} C_{\mathrm{abs}}^{\mathrm{PAH}}(a,\lambda)+(1-\xi_{\mathrm{PAH}})C_{\mathrm{abs}}^{\mathrm{gra}}(a,\lambda)
\end{equation}
where $C_{\mathrm{abs}}^{\mathrm{carb}}(a,\lambda)$ is the absorption cross section of the carbonaceous dust, $C_{\mathrm{abs}}^{\mathrm{PAH}}(a,\lambda)$ and $C_{\mathrm{abs}}^{\mathrm{gra}}(a,\lambda)$ the absorption cross sections of the PAHs and quantum-heated graphite grains respectively, $\xi_{\mathrm{PAH}}$ is the mixing ratio as defined by \cite{Li01}, and $0\leq\eta(a)\leq1$ is a function which parametrizes the PAHs destruction ratio, and that we will define below. It should be emphasized that this is not equivalent to modifying the grain sizes distribution function $\frac{1}{n_H}\frac{dn_{\mathrm{carb}}}{da}$ defined by \cite{Weingartner01} which would then impact the carbonaceous dust as a whole. PAHs are much more likely to be processed by UV radiation than quantum-heated graphite grains \citep[]{Luo92,Hunter01}, and we do not see any explicit reason to alter the latter category at first. This approach also agrees with the work of \cite{Compiegne08} who showed that the dust emission in the Horsehead Nebula photodissociation region appears to contain a smaller ratio of PAH vs. quantum-heated carbon grains than usual. 

\begin{figure}[htb!]
\centerline{\includegraphics[scale=0.37]{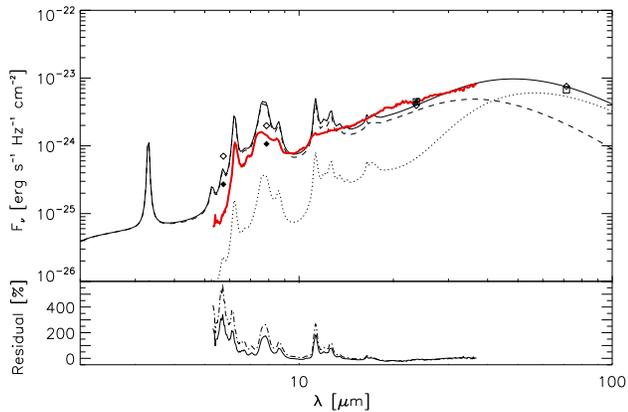}}
\caption{ Top : Same as Fig.~\ref{Fig:fit-1}, but for simulation SIM2 : 9\%/91\% e-mix with PAHs depletion as described in Eq.~\ref{Eq:xi-fit}. Bottom : fit residual relative to the recorded echo spectrum (full line) and the SIM1 residual (dash-dotted line) for comparison. PAHs depletion enables to slightly better fit the background level of the emission at small wavelengths without impacting the fit above $\sim$15~$\mu$m. }\label{Fig:fit-2}
\end{figure}

In Fig.~\ref{Fig:fit-2}, we present the result of the simulation labelled SIM2, with a 9\%/91\% e-mix, and for which we have taken the dust mix $\eta(a)$ as follow:

\begin{equation}\label{Eq:xi-fit}
\eta(a)=\begin{cases} 0 & \mbox{ if } a<5.5\textrm{\AA}\\
        0.5 & \mbox{ if } 5.5\le a<8.6\textrm{\AA}\\
        1 & \mbox{ if } a\ge 8.6\textrm{\AA} \end{cases}
\end{equation}
In other words, with $a=1.286\times N_{C}^{\frac{1}{3}}$ \citep{Li01}, we assume that there are no PAHs containing less than $\sim$80 carbon atoms, and a decrease of 50\% of the amount of PAHs containing from $\sim$80 to $\sim$300 carbon atoms, with respect to the initial ISM dust mix.  
 
As expected, the lack of small PAHs has an effect at small wavelengths, more precisely below $\sim$15~$\mu$m. The subsequent simulated IR spectrum has a trend that matches better the recorded echo spectrum. The 9~$\mu$m gap is better reproduced, and the 6.2~$\mu$m and 7.6~$\mu$m features are less intense than in SIM1. However, this update of the dust mix still cannot explain the small strength of the PAH features in the echo spectrum, a point which we will discuss in Sec.~\ref{Sec:dehydro}. The resulting total mass of echoing dust for this simulation is 7.0$\times$10$^{29}$~g, equivalent to a echo region density of $\rho_{\mathrm{echo}}$=679 H~cm$^{-3}$ or $1.3\times10^{-23}$~g~cm$^{-3}$, with a column density of $3.0\times10^{19}$~cm$^{-2}$. The UV component contributes 83.2\% to the total luminosity of the simulated echo in the 5-38 $\mu$m range.

The chosen form of $\eta(a)$ might appear somewhat artificial. It is important to clarify here that in our code, the UV burst spectrum and the $\eta(a)$ function are two different parameters treated separately in order to fit the recorded IR echo spectrum. A first justification for this particular $\eta(a)$ function would be to assume that the echoing dust content locally differs from the assumed \emph{standard} ISM dust distribution described previously, and that this precise echo region should not be considered to be pristine and unaffected by previous interactions prior to being heated by the Cas A SN light. For example, \cite{Micelotta10} showed that PAHs with less than 50 carbon atoms can be destroyed if encountering shocks with velocities greater than 100 km~s$^{-1}$. Furthermore, given that the \emph{standard} ISM dust content is not clearly defined, this justification represents a potential solution. 

A second explanation for the $\eta(a)$ function is to assume that the dust content is directly affected by the UV burst via photodissociation. In this case, the form of $\eta(a)$ is not independent but rather a direct consequence of the UV burst. In Sec.~\ref{Sec:UV-link}, we will discuss this possibility, and show that indeed, it is reasonable to assume PAH destruction leading to the form of $\eta(a)$ showed in Eq.~\ref{Eq:xi-fit} using a simple theoretical model of dust destruction.

In our simulations, we alter the dust composition of both the UV-heated and the optically-heated dust. As we will discuss in Sec.~\ref{Sec:UV-link}, the destruction of dust by the UV burst is not an equilibrium process counter balanced by dust production, but rather a instantaneous (on an ISM dust timescale) event. It appears rather unlikely that any dust formation process could replenish the PAHs destroyed by the UV burst before the optical component arrival. In that sense, it appears reasonable to assume that the dust composition for the optical slab has the same composition left-over from the UV burst. The study of other IR echo spectra should be, in the future, able to shine some more light on the behaviour of the dust after being processed by the UV burst of the SN.  

\subsection{The influence of PAH ionization}\label{Sec:ionization}

\begin{figure}[htb!]
\centerline{\includegraphics[scale=0.37]{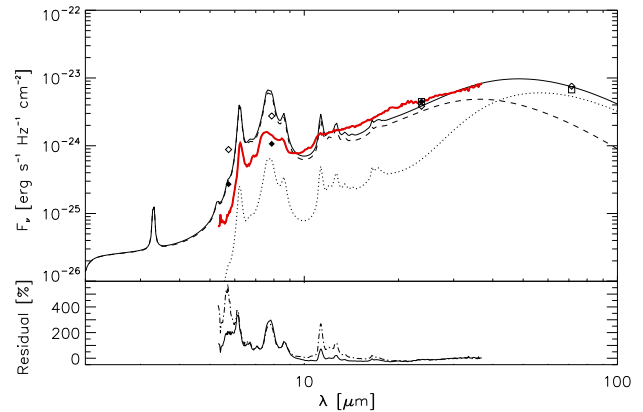}}
\caption{Same as Fig.~\ref{Fig:fit-2}, but for simulation SIM3: 9\%/91\% e-mix with PAHs depletion as described in Eq.~\ref{Eq:xi-fit}  and 100\% of PAHs ionized.}\label{Fig:fit-2b}
\end{figure}

As they have an ionization potential of the order of 7~eV \citep[][]{Li01}, the UV burst can ionize large amounts of PAHs in the echo region. Ionized PAHs are known \citep{Bakes01b} to have weaker 3.3~$\mu$m and 11~$\mu$m features compared to neutral PAHs. A stronger ionization fraction than in the normal ISM could be one explanation for the weak 11~$\mu$m complex. However, ionized PAHs also have stronger emission from 5-9~$\mu$m which would increase even more the discrepancies of the fit in this zone. Since we have adopted the ionization function of \cite{Li01} in our dust model ($\phi_{\mathrm{ion}}=\phi_{\mathrm{LD01}}$, see their Fig.~7), we can simulate the consequences of different ionization fractions. In SIM3, we have assumed all of the PAHs to be ionized ($\phi_{\mathrm{ion}}=1$), and kept all the other parameters identical to SIM2. The resulting fit to the recorded echo spectrum in shown in Fig.~\ref{Fig:fit-2b}. The resulting total mass of echoing dust is 7.0$\times$10$^{29}$~g, equivalent to an echo region density of $\rho_{\mathrm{echo}}$=679 H~cm$^{-3}$ or $1.3\times10^{-23}$~g~cm$^{-3}$, with a column density of $3.0\times10^{19}$~cm$^{-2}$.

As expected, increasing the ionization fraction increases the 6.2+7.7~$\mu$m to 11.3$\mu$m ratio of the PAHs features. As a consequence, the increased PAH ionization fraction improves the fit in the 11.3~$\mu$m region by reducing the intensity of the PAH features, but also significantly worsen the fit for the 6-8~$\mu$m region. It can also be noted than even with a 100\% of PAHs being ionized, the 11.3~$\mu$m feature remains strong - too strong with respect to the recorded echo spectrum. Hence, if PAHs are in principle likely to be strongly ionized by the passage of the UV burst, the extent to which ionization influence the shape of echo spectrum is not evident, and another mechanism appears to be required to explain the very weak PAH features. In the remaining simulations of this paper, we will assume the original ionization fraction $\phi_{\mathrm{ion}}=\phi_{\mathrm{LD01}}$ because, from our simulations, the role played by PAHs ionization is not clear at this stage.

\subsection{The weak 11~$\mu$m complex: a signature of dehydrogenation}\label{Sec:dehydro}

The lack of any reasonably strong features in the 11~$\mu$m region in the recorded IR echo spectrum is quite puzzling. These features are usually associated with C-H out-of-plane bending modes \citep{Bakes01,Hony01}. Dehydrogenation might provide an alternative explanation to ionization for this peculiar 11~$\mu$m complex. 

Dehydrogenation occurs when a PAH loses some H atoms, in our case by absorption of a UV photon. Looking closer to the recorded spectrum, one can note, along with a weak 11~$\mu$m complex, very little emission around the 5.27~$\mu$m and 5.7~$\mu$m features (C-H bend/stretch combination mode, see \cite{Bakes01, Draine07}). Those characteristics of the recorded echo spectrum point towards a possible depletion of emission originating from C-H bonds, and suggest a strongly dehydrogenated state of the PAHs. If the intensity of the 11~$\mu$m feature suggest a rather high dehydrogenation ratio, the presence of the 8.6~$\mu$m C-H in plane bending is an indicator that not all hydrogen atoms have disappeared. 

To take dehydrogenation of PAHs into account in our simulations, we define the number of hydrogen atoms in a dehydrogenated PAH $N_{H}^{*}$ :
\begin{equation}\label{Eq:dehydro}
N_{H}^{*}=\delta_{\mathrm{hydro}}\times N_{H}
\end{equation}
with $\delta_{\mathrm{hydro}}$ the dehydrogenation ratio, and $N_{H}$ the number of hydrogen atoms in the \emph{un-altered} PAH molecule containing $N_{C}$ carbon atoms.

\begin{figure}[htb!]
\centerline{\includegraphics[scale=0.37]{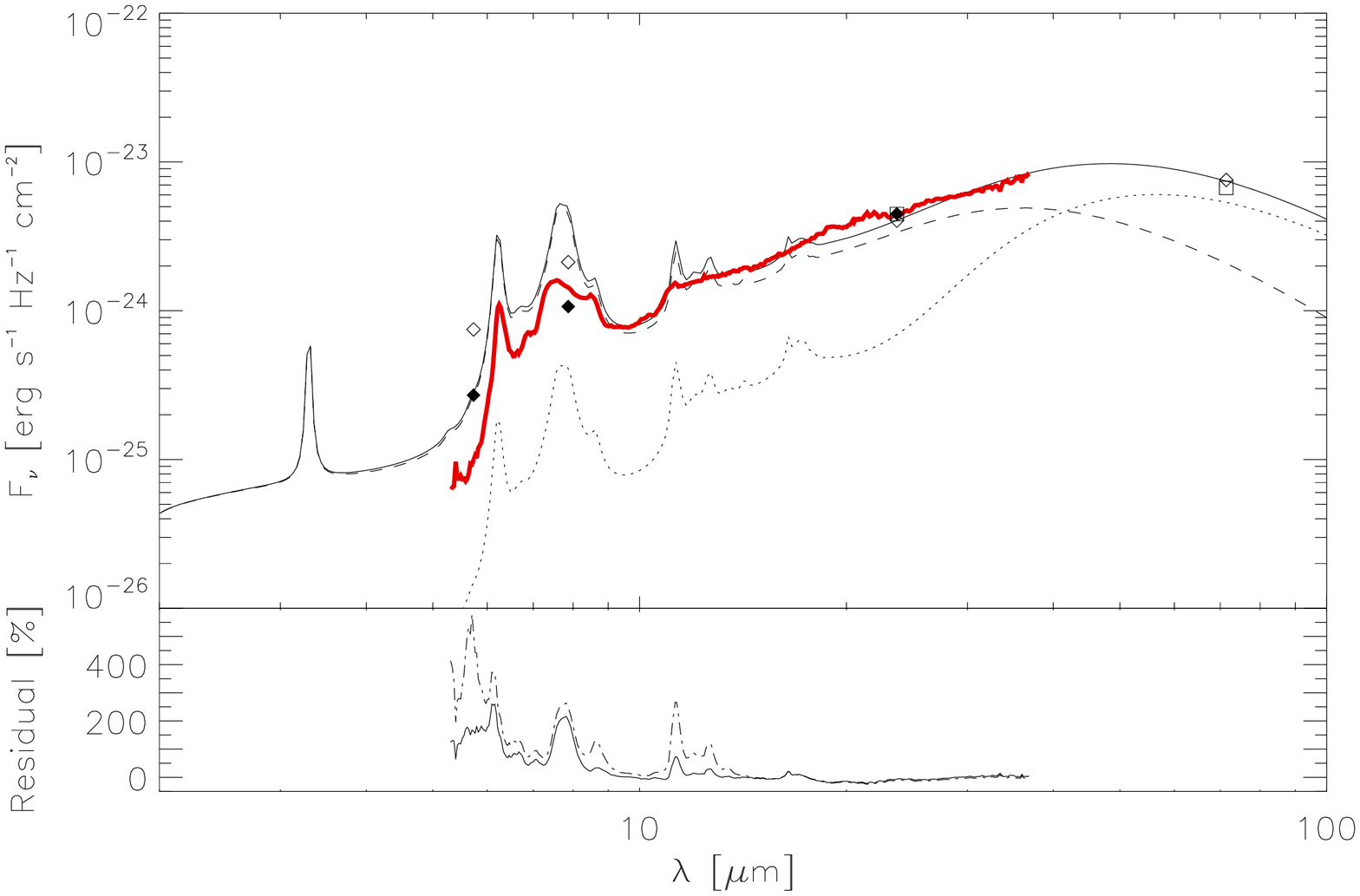}}
\centerline{\includegraphics[scale=0.37]{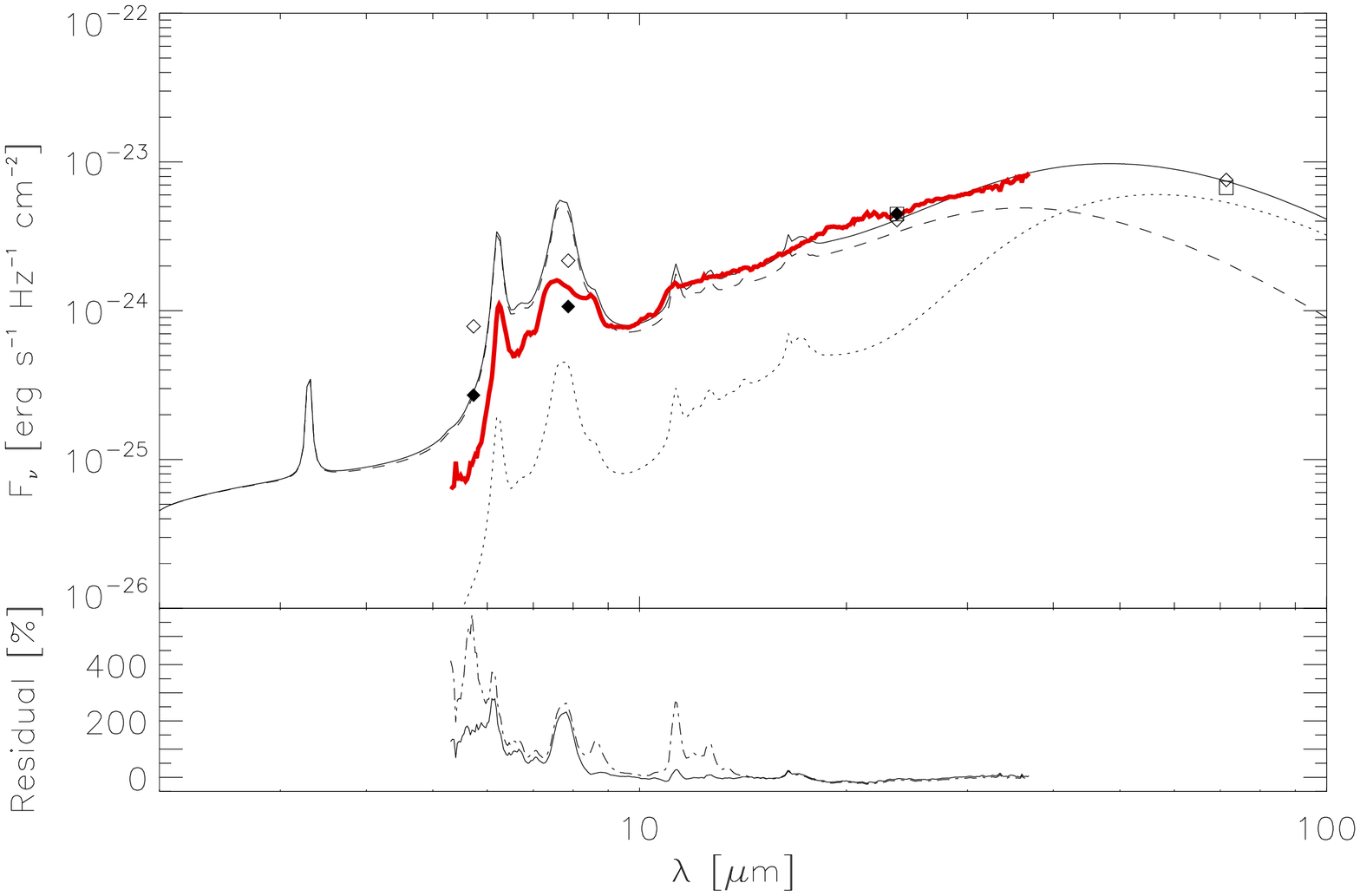}}
\centerline{\includegraphics[scale=0.37]{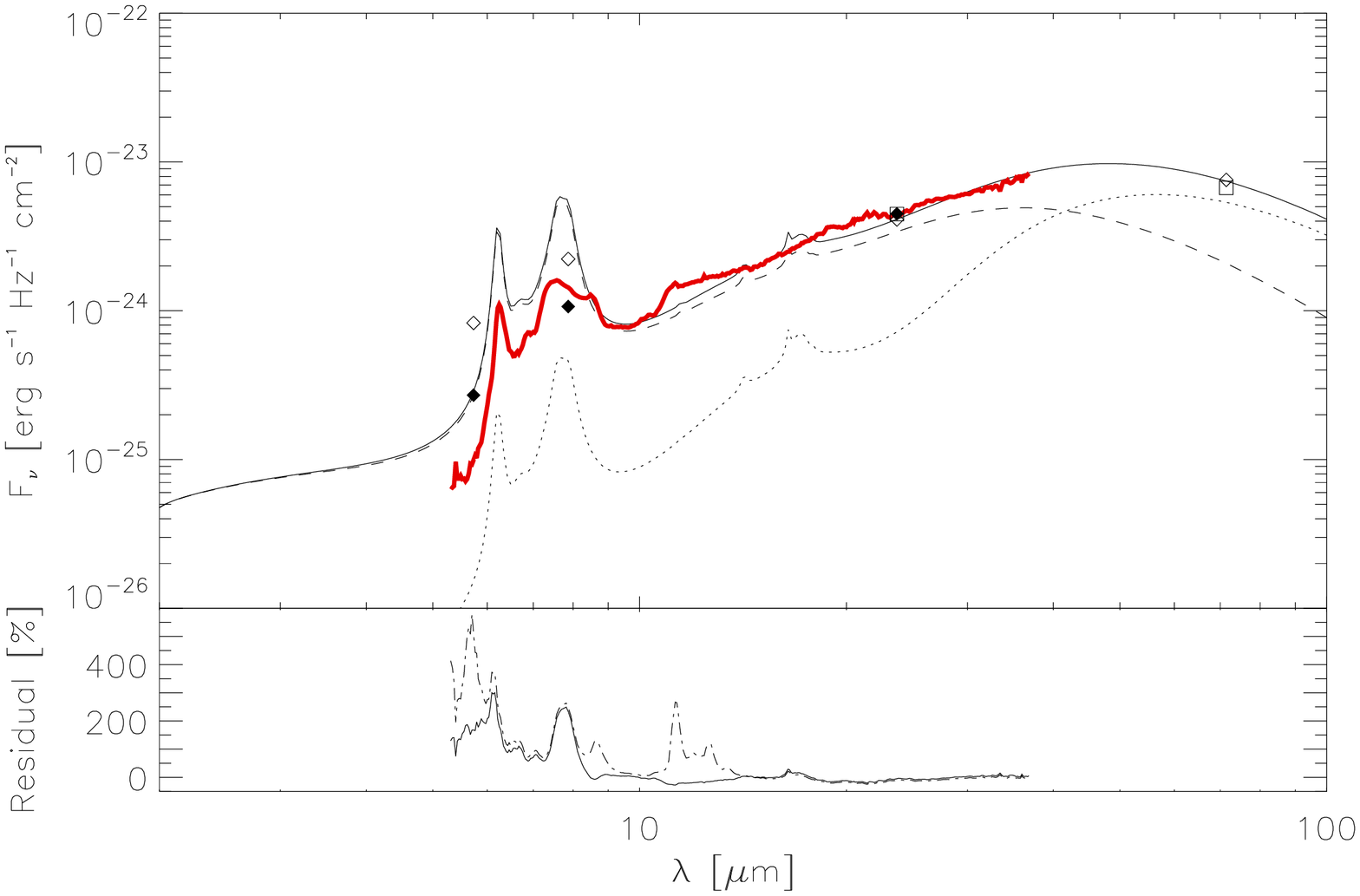}}
\caption{Same as Fig.~\ref{Fig:fit-2}, but for simulations SIM4: 9\%/91\% e-mix with PAHs depletion as described in Eq.~\ref{Eq:xi-fit}  and PAHs dehydrogenated at 60\% (top), 80\% (center) and 100\%  (bottom). Dehydrogenation might be a key factor able to explain the very weak 11~$\mu$m PAH complex.}\label{Fig:fit-3}
\end{figure}

In Fig.~\ref{Fig:fit-3}, we show the simulated spectra SIM4 where the amount of H atoms per PAH is 40\% (top), 20\% (center) and 0\% of the original value (i.e. $\delta_{\mathrm{hydro}}=0.4;0.2;0.0$)). All the other parameters are identical to the simulation SIM2 shown in Fig.~\ref{Fig:fit-2}. For the three cases, the resulting total mass of echoing dust is 7.0$\times$10$^{29}$~g, equivalent to an echo region density of $\rho_{\mathrm{echo}}$=679 H~cm$^{-3}$ or $1.3\times10^{-23}$~g~cm$^{-3}$, with a column density of $3.0\times10^{19}$~cm$^{-2}$. Note that we have assumed that the 5.2~$\mu$m and 5.7~$\mu$m features are functions of the hydrogenation ratio H/C, thus departing on this aspect from \cite{Draine07}. It is not clear why they did not have their intensities as a function of the H/C ratio, but given our recorded IR echo spectrum, we believe this update to be reasonable. The UV component of the simulated echo spectrum accounts for 83.4\%, 83.3\% and 83.3\% of the total luminosity of the simulated echo for the 40\%, 20\% and 0\% hydrogenation simulations. The respective parameters of all our simulations, presented in Fig.~\ref{Fig:fit-1} to Fig.~\ref{Fig:fit-3}, are shown in Table~\ref{Table:param}.

\begin{table*}[htb!]
\begin{center}\caption{Various simulations and associated parameters.}\label{Table:param}

\begin{tabular}{c c c c c p{1.2cm} p{1.2cm} p{2cm} p{3cm}}
&&&&\\
\tableline
\tableline
Name& Emissivity mix & Dust mix &$\phi_{\mathrm{ion}}$& $\delta_{\mathrm{hydro}}$ \tablenotemark{a}  & \centerline{$M_{\mathrm{dust}}$} \centerline{$10^{29}$ [g]} &\centerline{ $\rho_{\mathrm{echo}}$} \centerline{[H~cm$^{-3}$]}&\centerline{Echo region} \centerline{column density} \centerline{$10^{19}$[~H~cm$^{-2}$]}&\centerline{UV/opt. contribution} \centerline{to the total luminosity}\\
\hline
SIM1&9\%~UV + 91\%~opt.&ISM&$\phi_{\mathrm{LD01}}$&0&$\centerline{6.6}$& $\centerline{640}$&\centerline{2.8}&\centerline{80.0\% / 20.0\%}\\
SIM2&9\%~UV + 91\%~opt.&ISM $\times\eta(a)$ \tablenotemark{b}&$\phi_{\mathrm{LD01}}$&0&$\centerline{7.0}$& \centerline{679}&\centerline{3.0} &\centerline{83.2\% / 16.8\%}\\
SIM3&9\%~UV + 91\%~opt.&ISM$\times\eta(a)$ \tablenotemark{b}&100 \%&0&$\centerline{7.0}$& \centerline{679}&\centerline{3.0} &\centerline{83.8\% / 16.2\%}\\
SIM4&9\%~UV + 91\%~opt.&ISM$\times\eta(a)$ \tablenotemark{b}& $\phi_{\mathrm{LD01}}$&0.4/0.2/0.0&$\centerline{7.0}$& \centerline{679}&\centerline{3.0}&\centerline{83.4\% / 16.6\%}\\
\tableline
\end{tabular}
\tablenotetext{1}{As defined in Eq.~\ref{Eq:dehydro}.}
\tablenotetext{2}{ $\eta(a)$ as defined in Eq.~\ref{Eq:xi-fit}.}
\end{center}
\end{table*}

The 11 $\mu$m region is best fitted by the 80\% dehydrogenation curve. Our aim in SIM4 was mainly to highlight the effect of dehydrogenation on the simulated echo spectrum. We are here, quite clearly, pushing our simulations to their limits. In reality, ionization of dust might play along with dehydrogenation and reduce this value. It will likely be very interesting to study how universal these C-H features strengths are in various light-echo spectra, and if they display any spatial variations, change with the SN type or are function of the echo region-to-SN distance, for example. We note here that in our simulation, dehydrogenation seems to be more important (and plays a bigger role) than PAHs destruction - a rather intuitive fact. One would expect PAHs first to get dehydrogenated by loosing their surrounding H-atoms, and only then, if the radiation field is strong enough, to get destroyed via further photodissociation events inside their more robust carbon skeleton.

It can be noted that compared to the SIM2 simulation, the 6-9~$\mu$m PAH features are slightly stronger in the SIM3 simulation. In fact, looking closely at the residual curves, the stronger the dehydrogenation value, the stronger the 6-9~$\mu$m features get. This fact most likely arises as we alter and artificially remove hydrogen atoms from PAHs, which will re-radiate more of their internal energy using C-C bonds. 

At this point, we are then able to reproduce closely the recorded echo spectrum and photometric measurements from 5 to 70~$\mu$m, with the exception of the 5-9~$\mu$m region, in which our simulation has too strong PAHs features. We do not have direct explanation for this overestimate, however, it appears quite clear that PAHs are strongly altered in this echo region. Especially, a more careful treatment of PAHs dehydrogenation than the one adopted here would be required to better understand the consequences of a strong and short-lived UV burst, and might potentially be able to explain the peculiar 5-9~$\mu$m region.

Finally, we note that based on the classification of \cite{Peeters02}, our recorded echo spectrum falls into the so-called AB' category, based on its 7.7~$\mu$m complex. This is a peculiar category containing only two objects, a non-isolated Herbig~Ae~Be star and a post-AGB star, among the 57 sources studied, for which the 7.7~$\mu$m complex is actually composed of two features at 7.6~$\mu$m and 7.8~$\mu$m with roughly equal strengths. This association of the unusual light-echo spectrum with the peculiar AB' class might eventually turn out to be of interest when trying to understand the origin of the 7.8~$\mu$m feature \citep{Peeters02}.

\section{Comparison with previous work}\label{Sec:dwek}

\begin{figure}[htb!]
\centerline{\includegraphics[scale=0.37]{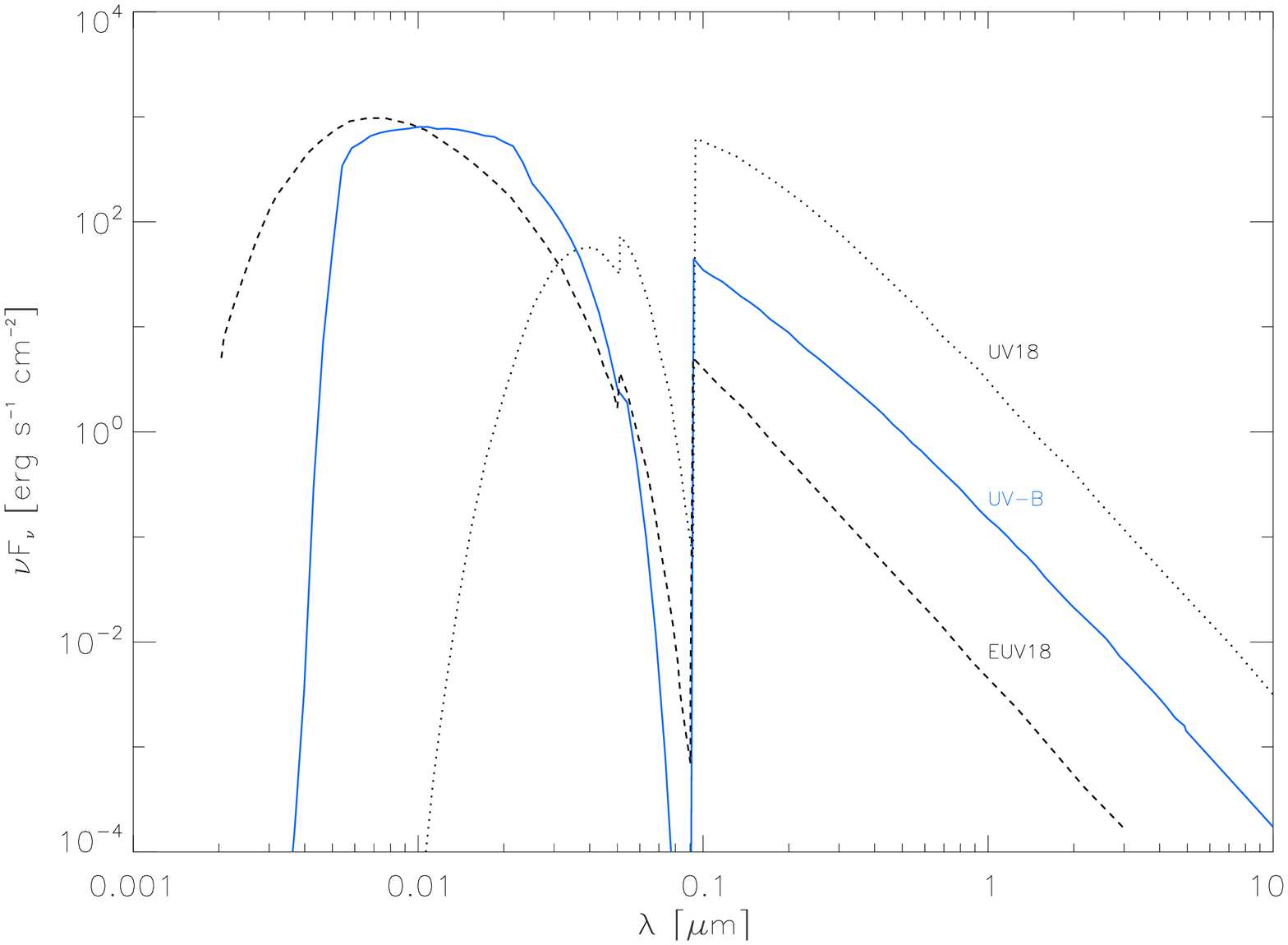}}
\caption{Comparison of our adopted UV burst spectrum (UV-B, full line, in blue in the online version) with the UV18 (dotted line) and EUV18 (dashed line) spectra of \cite{Dwek08} (see their Fig.~6). The UV18 and EUV18 curves have both been scaled to account for the distance difference of their associated echo (160 ly vs 199 ly). We have taken their luminosity to be $0.11\times10^{12}$ L$_{\sun}$ and $0.15\times10^{12}$ L$_{\sun}$ respectively, as mentioned by \cite{Dwek08} (see their Table~3). }\label{Fig:dwek}
\end{figure}

Using a similar approach to the present work, \cite{Dwek08} presented simulations of several Cas-A light echoes in the 10-40~$\mu$m range. They especially  experimented with three different types of heating spectra : optical light curves, UV bursts, and (so-called) extreme  UV bursts (or EUV). Because they used several different bursts characteristics, spanning a range of luminosities and temperatures, and managed to successfully reproduce their echo spectra, it is interesting to compare our UV-B burst based on our knowledge of the Cas A SN type to their models. In Fig.~\ref{Fig:dwek}, we plot our UV-B burst against their 'UV18' and 'EUV18' curves. Both are black-body spectrum with a temperature T$_{\mathrm{uv18}}=5\times10^4$~K and T$_{\mathrm{euv18}}=5\times10^5$~K. The number \emph{18} refers to the assumed SN-echo column density, $1.5\times10^{18}$~cm$^{-2}$. Note that compared to their Fig.~6, both UV18 and EUV18 curves have been scaled to account for the different distance to the light echo (160 ly versus 199 ly), as well as by 0.11 and 0.15 respectively to match the averaged luminosity over all their echo regions for the respective burst model (see Table~3 in \cite{Dwek08}). 

It is interesting to note that our respective burst intensities agree very well. The total luminosities of L$_{uv18}=4.2\times10^{44}$~erg~s$^{-1}$ and L$_{euv18}=5.7\times10^{44}$~erg~s$^{-1}$ are slightly lower but nevertheless in good agreement with our UV-B luminosity of L$_{uv-B}=6.4\times10^{44}$~erg~s$^{-1}$. The difference is mostly a consequence of the shape modification of our UV-B burst. The luminosity of our black-body UV burst prior to the modification is $5.9\times10^{44}$~erg~s$^{-1}$, in perfect agreement with the work of \cite{Dwek08}.
 
Because they did not use data below 13~$\mu$m, \cite{Dwek08} were not able to find a model fitting the data significantly better than the others, with exception to the optical ones that proved not to be able to reproduce the light echo shape. Our simulations confirm these conclusions, in that the 10-40~$\mu$m range in a light echo spectrum is mostly influenced by the UV burst heating the dust. Distinguishing between various UV burst characteristics only becomes possible at smaller wavelengths, below 10~$\mu$m, where the behaviour of the various PAH features, as well as the global shape of the spectrum contains unique signatures of the intensity and temperature of the UV burst. In that sense, our simulations suggest the EUV models of \cite{Dwek08} to be more likely, especially as energetic photons are required to explain the weak 11~$\mu$m PAH complex with dehydrogenation. As for the optical light curve, accounting for a mere 17\% in the 4-38~$\mu$m range in our simulations, its influence really becomes critical above 50~$\mu$m, explaining why \cite{Dwek08}, which did not use 70~$\mu$m photometry data, disregarded its influence in their fit.

In our various simulations, we find an echo region density of 640-679 H~cm$^{-3}$, relatively comparable to the value of $\sim$385 H~cm$^{-2}$ of \cite{Dwek08}. Note that these do not, in principle, need to agree, as different light echoes in different locations have been considered in both works. In terms of column density, we find a much higher value, roughly $3.0\times10^{19}$~H~cm$^{-2}$, as compared to $5\times10^{17}$~H~cm$^{-2}$. This difference arises as we model our echo region as containing both a UV-slab and an optical slab. Taking only the UV-component of our echo model into account to fit the recorded spectrum, we would find a dust mass of $4.77\times10^{28}$ g. Subsequently, only taking the UV- slab as defining the volume of the echo region, our simulations would lead to a column density of $2.0\times10^{18}$~H~cm$^{-2}$, more consistent with the results of \cite{Dwek08}, especially their Echo~6 (see their Table~4).

\section{Destroying PAHs}\label{Sec:UV-link}
In order to improve the fit of the recorded spectrum (especially the zone below 15~$\mu$m), we have updated the dust mix in our simulations using the $\eta(a)$ factor (see Eq.~\ref{Eq:xi}). The form of $\eta(a)$ as defined in Eq.~\ref{Eq:xi-fit} has been plugged in as an individual parameter of the simulation. However, if this lack of dust is linked to the UV burst, then $\eta(a)$ is not independent. The strength of the burst will determine the amount and sizes of PAHs destroyed, while the duration of the burst will determine the total column density of the echo region in which the PAHs are destroyed, and the relative mix of UV-heated and optically-heated grains. This dependency of $\eta(a)$ can be used as a consistency check for our PAHs destruction hypothesis. 

The fate of PAHs in the ISM, and the influence of photodissociation, ionization and collisional processes has been subject to several studies \citep[e.g.][]{LePage01, Micelotta10b}. Here, we consider dust destruction using the same approach as \cite{Siebenmorgen10}. Let us define F$_{\mathrm{burst}}$($\nu$), the incoming radiation flux heating the dust ; C$_{\mathrm{abs}}$($\nu$,~N$_C$), the PAH absorption cross section as a function of the frequency $\nu$ and the number of C-atoms N$_C$. Excited PAHs will de-excite themselves by IR emission in a time t$\sim$1~s, unless their internal energy is higher than a threshold value $E_{\mathrm{lim}}$. In this case, the PAH molecule will be disrupted and loose one or more atoms. For a PAH with $N_C$ carbon atoms, we take the threshold energy input within 1~s for PAH destruction, $E_{\mathrm{lim}}$, to be :
\begin{equation}
\frac{E_{\mathrm{lim}}}{[\mathrm{eV}]}=\frac{N_C}{2} 
\end{equation}
This relation is based on the assumption that :
\begin{equation}\label{Eq:elim}
E_{\mathrm{lim}}=3N_{C}kT_{\mathrm{dis}}\sim0.1N_{C}E_0
\end{equation}
with $T_{dis}$ the minimum temperature for destruction, and $E_0=5$~eV the critical (Arrhenius) energy of the atom being ripped off the molecule. Let us subsequently define $x$ the number of dissociated atoms in a PAH of size $N_C$ receiving an energy $E_{\mathrm{in}}\geq E_{\mathrm{lim}}$ within 1~s of time, and $E_{\mathrm{kin}}$ the kinetic energy of one ripped off atom. Then, similarly to Eq.~\ref{Eq:elim} :
\begin{equation}
E_{\mathrm{in}}-x(E_0+E_{\mathrm{kin}})=3(N_c-x)kT_{\mathrm{dis}}
\end{equation}
In words, as a PAH is heated by an incoming energy $E_{\mathrm{in}}$ higher than the threshold value $E_{\mathrm{lim}}$, it will loose $x$ atoms until its internal temperature decreases down to $T_{dis}$. Rearranging the terms and using Eq.~\ref{Eq:elim}, it is possible to express $x$ as a function of the incoming energy $E_{\mathrm{in}}$ and the PAH size N$_C$ :
\begin{equation}\label{Eq:x}
x=\frac{E_{\mathrm{in}}}{5 [\mathrm{eV}]}-\frac{N_C}{10}
\end{equation}
where $E_{\mathrm{in}}$ can be reached by the absorption of one single energetic photon, or several less energetic photons. We here strictly follow the work of \cite{Siebenmorgen10}, which we refer the reader to for more details. 

\begin{figure}[htb!]
\centerline{\includegraphics[scale=0.52]{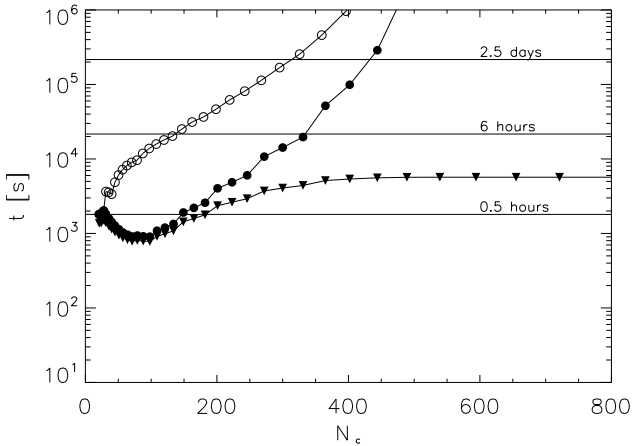}}
\caption{Mean time t required for a PAH particle to get hit by one destructive photon (filled-dots line) and to be completely destroyed (empty-dots line) as a function of the number of carbon atoms $N_C$ in the PAH. The triangle-symbols line shows the time required to get hit by a photon of any energy, including those not able to rip off atoms of the PAH structure.}\label{Fig:tin}
\end{figure}

At the estimated distance of the echo region, the amount of time required to get hit by a photon (triangle-symbols line in Fig.~\ref{Fig:tin}), assuming the template UV incoming radiation field described in Sect.~\ref{Sec:SNbursts}, is more than one second for PAHs with $N_{C}\leq 10^{5}$. Hence, those PAH can only loose atoms if they are hit by one single photon with an energy $h\nu\geq E_{\mathrm{lim}}$. 

PAHs with sizes above 10$^{5}$~$N_C$ will, in average, get hit by several photons every second, however, taking into account the mean energy of the absorbed photons, the total absorbed energy will not reach the limit value $E_{\mathrm{lim}}$ for the radiation field strengths considered in this work. The optical light curve, as considered in this study, does not contain photons energetic enough to be able to destroy PAHs. 

For PAHs with $N_C\leq 10^5$, the absorbed \emph{destroying} energy $\dot{E}_{\mathrm{in,}h\nu>E_{\mathrm{lim}}}$ per PAH per second as a function of the PAH size is :
\begin{equation}
\dot{E}_{\mathrm{in,}h\nu\geq E_{\mathrm{lim}}}(N_C)=\int_{h\nu=E_{\mathrm{lim}}}^{\infty} F_{\mathrm{burst}}(\nu)C_{\mathrm{abs}}(\nu,N_C)d\nu
\end{equation}
This corresponds to a number of \emph{destroying} photons $\dot{N}_{\gamma\mathrm{,in,}h\nu>E_{\mathrm{lim}}}$ absorbed per PAH per s :
\begin{equation}
\dot{N}_{\gamma\mathrm{,in,}h\nu\geq E_{\mathrm{lim}}}(N_C)=\int_{h\nu=E_{\mathrm{lim}}}^{\infty} d\nu\frac{ F_{\mathrm{burst}}(\nu)C_{\mathrm{abs}}(\nu,N_C)}{h\nu}
\end{equation}
which finally leads to the mean time $t_{\mathrm{in,}h\nu\geq E_{\mathrm{lim}}}$ necessary to absorb one \emph{destroying} photon :
\begin{equation}\label{Eq:tin}
t_{in,h\nu\geq E_{\mathrm{lim}}}(N_C)=\frac{1}{\dot{N}_{\gamma\mathrm{,in,}h\nu\geq E_{\mathrm{lim}}}(N_C)}
\end{equation}
and the mean energy of the destroying photons :
\begin{equation}\label{Eq:numean}
h\bar{\nu}_{\mathrm{in,}h\nu\geq E_{\mathrm{lim}}}(N_C)=\frac{\dot{E}_{\mathrm{in,}h\nu\geq E_{\mathrm{lim}}}(N_C)}{\dot{N}_{\gamma\mathrm{,in,}h\nu\geq E_{\mathrm{lim}}}(N_C)}
\end{equation} 

In Fig.~\ref{Fig:tin}, we show the time required by a PAH to absorb one destructive photon, t$_{\mathrm{in,}h\nu\geq E_{\mathrm{lim}}}$, as a function of PAH size $N_C$ (filled-dots line), along with the average time for a PAH to get hit by a photon of any energy (triangle-symbols line). We use our SN burst templates described in Sec.~\ref{Sec:SNbursts} and assume the distance of 199 light years (see Sec.~\ref{Sec:obs}) from Cas A to the echo region. Under those assumptions, we see that it takes $\sim$2.5 days for a PAH with $N_C$=450 C-atoms to be hit by one destructive photon with the UV burst described in Sect.~\ref{Sec:SNbursts} shining on the dust particles.

We also plot in Fig.~\ref{Fig:tin} a higher-end estimate of the time required for a PAHs of a certain size to be destroyed via photodissociation (empty-dots line), based on the conservative following approach: assuming an initial PAH size $N_{C,0}$, we find the mean time required to get hit by a destructive photon (Eq.~\ref{Eq:tin}), calculate the number $x$ of atoms kicked off (via Eq.~\ref{Eq:x}) given the destructive photon mean energy (from Eq.\ref{Eq:numean}). We then start again with the new, smaller, PAH with $N_{C,1}=N_{C,0}-x$ atoms, and iterate the process until we reach $N_{C,n}\leq20$, a value below which the PAHs are photolitically unstable \citep{Li01,LePage03}. Summing the time of the various steps of the loop gives us the mean destruction time $t_{\mathrm{destruct}}$ for a PAH of size $N_{C,0}$. This time can be considered as an over estimate, as a PAH, after having lost some atoms, might not turn into a stable PAHs of the new size. The C-atoms will likely be ripped off at random position, and thus weaken the PAH structure. In that sense, it may be possible that less impacts of destroying photons are required in order to achieve the destruction of the PAH - this, however, is not certain \citep[see][]{LePage03}.

As seen in Fig.~\ref{Fig:tin}, the destruction time increases rapidly with the PAH size, and it already takes $\sim$2.5 days in order to destroy PAHs with an initial size of $\sim$300 C-atoms. Let us recall here that assuming all PAHs below 80 carbon atoms and 50\% of PAHs from 80 to 300 carbon atoms to be destroyed lead to the SIM2 simulation. These very similar values suggest, under the assumptions of this simple PAH destruction model, that the destruction of PAHs by the UV burst appears plausible and logical. It should be noted here that we have neglected any dust processing mechanisms other than photodissociation, such as recombination, hydrogenation or the coagulation of small molecules on bigger ones \citep[]{Hunter01}. We do not expect these to have an impact as important as photodissociation in shaping the echo spectrum, as the reaction timescale are expected to be significantly larger than the duration of the UV-burst, typically of the order of a few years for hydrogenation and recombination \citep[in a standard ISM environment, see for example][]{LePage01,LePage03}.

Such a dust destruction scenario also appears of special interest when considering the fact that \cite{Krause08} have found strong [C~I]$\lambda$9850~{\AA}  carbon emission in their optical spectrum of an echo region around Cas A. The [C~I] lines can be excited via two main mechanisms: radiative recombination and collisional excitation by electrons. \cite{Escalante91} suggested the first mechanism to explain the [C~I] emission lines from M~42 and NGC~2024, requiring strong radiation (10$^3$ to 10$^6$ times the ISRF) heating gas with a density of 10$^5$~H~cm$^{-3}$; a radiation field intensity not so different from the one considered in this work. In our dust model, the C/H ratio in PAHs is 16.1~ppm for N$_C<$100, 26.1~ppm for N$_C<$200 and 36.1~ppm for N$_C<$500 (see Table~3 in \cite{Draine07}). Dust destruction in the form of $\eta(a)$ described in Eq.~\ref{Eq:xi} will therefore, if affected PAHs are entirely disrupted, increase the ISM gas phase abundance of C atoms by $\sim$24~ppm. Standard carbon abundances in the gas phase of low to moderate density ISM have been measured to be of the order of 150 ppm \citep[e.g.][]{Cardelli96,Sofia97}, while \cite{Dwek97} found abundances of the order of 50-100~ppm in the Cold Neutral Medium. Recently, using a different estimation method based on the strong transition of CII~$\lambda$1334~{\AA}, \cite{Sofia11} suggested that the carbon gas phase abundance might be lower by $\sim$40\% than previous values found using methods based on a weak intersystem absorption transition. In any case, our dust destruction scenario could increase the amount of C atoms in the gas phase by as much as 10\%-20\%, which could in turn potentially affect the [C~I] line by the same factor. If this suggest that the [C~I]]$\lambda$9850~{\AA} detected by \cite{Krause08} is not an unambiguous and direct confirmation of our scenario of PAHs destruction, it can not be ruled out that destroyed PAHs influence its strength. Clearly, the study of other optical light echoes is required to better understand the origin and the ubiquity (or not) of the [C~I]$\lambda$9850~{\AA} line.

\section{Conclusion}\label{Sec:conclusion}
We have obtained a Spitzer IRS spectrum and MIPS photometric measurements of an IR echo region around the Cas A supernova. Using simulations of an ISM dust mix containing both carbon and silicate quantum-heated and thermal grain as well as PAHs, we can, to a large extent, reproduce the recorded IR  echo spectrum. Specifically :
\begin{enumerate}
\item We find that the infrared echo spectrum can be described as the thermal signature from an UV and an optical component. Those UV and optical component characteristics correspond to the values of the type IIb SN 1993J which is considered to be representative of the Cas A supernova based on its optical spectrum. Their respective contribution in our best fit to the echo total luminosity are $\sim$83\% and $\sim$17\% (in the 5-38 $\mu$m range).
\item The influence of the optically-heated dust becomes more important than UV-heated dust above $\sim$50$\mu$m. Especially, the presence of the optical component in our heating spectrum is consistent with the 70$\mu$m photometric measurement of the recorded echo.
\item The spectrum below 15~$\mu$m is better reproduced when removing artificially small PAHs with less than 80 carbon atoms, and by removing 50\% of PAHs with 80 to 300 carbon atoms.
\item The especially weak 11$\mu$m complex of PAH features in the recorded echo spectrum can be reproduced when introducing 80\% of dehydrogenation for all PAHs in our simulation. 
\end{enumerate}
Our best fit simulation implies a lower limit for the density of this echo region of $\rho_{\mathrm{echo}}=$679~H~cm$^{-3}$. The removal of small PAHs is consistent with a simple and rather conservative theoretical model of PAHs destruction via photodissociation assuming the considered UV burst radiation field. The strong dehydrogenation ratio of PAHs, as compared to the amount of PAHs destruction, is also consistent with this picture, where PAHs tend to loose hydrogen atoms first.  In the future, a more careful modelling of PAHs destruction and dehydrogenation as compared to the one adopted here could lead to an even better fit to the recorded IR echo spectrum. The relatively weak 5-9~$\mu$m PAH complex, where our simulations differ from the recorded echo spectrum by a factor of $\sim$250\%, remains to be explained.

The approach we adopted in this work, in order to study dust processing in the ISM, has a great potential. Our results aim at being tested towards other echo regions around Cas A and other SNR, and this preliminary study shows the wealth of potential outcomes. The unique nature of IR echoes enables the study of otherwise cold and very faint ISM structures being strongly processed over a very short period of time. Such ISM clouds, potentially being initially pristine, undisturbed, and located away from any altering radiation sources, can provide unique insights in the ISM physics. Furthermore, the large amount of echoes provides us with an extensive number of locations and environments to test the simulations and conclusions presented in this article. It is currently rather difficult to obtain near IR spectra of echoing regions, now that the Spitzer Space Telescope has reached the end of its duty cycle. Fortunately, there are some exciting prospects for observations at those wavelengths, such as the \emph{James Webb Space Telescope} (JWST) \citep{Gardner06} or the \emph{Stratospheric Observatory for Infrared Astronomy} (SOFIA) \citep{Becklin07,Gehrz09} which has seen its first light in May 2010. These telescopes and their instruments will be ideally suited to record infrared spectra of echoing regions, and we expect that infrared echoes will provide many new and unique insights into the ISM dust composition and chemistry, as well as into the SN symmetries and characterization.

\acknowledgments
We thank the anonymous referee for his/her constructive review. Most of this work has been carried out during two visits of F.V. at the Max-Planck-Institut f$\ddot{\textrm{u}}$r Astronomie in Heidelberg. F.V. thanks Alexandra Bohm for her help dealing with the logistic related to those visits. This research has made use of NASA's Astrophysics Data System.

\end{document}